\documentclass[11pt]{article}
\usepackage[margin=1in]{geometry}
\usepackage{setspace}
\usepackage{enumitem}
\usepackage{float}
\usepackage{amsmath}
\usepackage{amsfonts}
\usepackage[none]{hyphenat}
\usepackage{amssymb}
\usepackage{latexsym}
\usepackage{graphicx}
\usepackage[english]{babel}
\usepackage{cite}
\usepackage{tocloft}
\usepackage{tikz}
\usepackage{hyperref}
\hypersetup{
    colorlinks=true,
    linkcolor=blue, 
    citecolor=orange, 
    urlcolor=teal   
}
\begin{document}
\input epsf

\def\p{\partial}
\def\h{{1\over 2}}
\def\be{\begin{equation}}
\def\bea{\begin{eqnarray}}
\def\ee{\end{equation}}
\def\eea{\end{eqnarray}}
\def\d{\partial}
\def\la{\lambda}
\def\eps{\epsilon}
\def\bb{\bigskip}
\def\mm{\medskip}
\def\beq{\begin{equation}}
\def\eeq{\end{equation}}
\newcommand{\dm}{\begin{displaymath}}
\newcommand{\edm}{\end{displaymath}}
\renewcommand{\b}{\tilde{B}}
\newcommand{\gm}{\Gamma}
\newcommand{\ac}[2]{\ensuremath{\{ #1, #2 \}}}
\renewcommand{\ell}{l}
\newcommand{\z}{\ell}
\newcommand{\newsection}[1]{\section{#1} \setcounter{equation}{0}}
\def\bb{$\bullet$}
\def\Qbar{{\bar Q}_1}
\def\QPbar{{\bar Q}_p}
\def\p{\partial}
\def\h{{1\over 2}}
\def\be{\begin{equation}}
\def\bea{\begin{eqnarray}}
\def\ee{\end{equation}}
\def\eea{\end{eqnarray}}
\def\b{\bigskip}
\def\r{\rightarrow}
\def\t{\tilde}
\def\nn{\nonumber\\}
\def\l{$\Lambda$~}
\def\hbar{{\mathchar'26\mkern-9muh}}
\def\MM#1{{\textcolor{red}{[MM: #1]}}}

\def\q{\quad}

\def\bn{B_\circ}

\def\MM#1{{\bf \textcolor{red}{MM:} }{\textcolor{blue}{#1}}}
\def\h{{1\over 2}}
\def\t{\tilde}
\def\r{\rightarrow}
\def\nn{\nonumber\\}
\let\bm=\bibitem
\def\Kt{{\tilde K}}
\def\b{\bigskip}

\let\p=\partial

\newcommand\blfootnote[1]{%
  \begingroup
  \renewcommand\thefootnote{}\footnote{#1}%
  \addtocounter{footnote}{-1}%
  \endgroup
}
\numberwithin{equation}{section}
\newcounter{daggerfootnote}
\newcommand*{\daggerfootnote}[1]{%
    \setcounter{daggerfootnote}{\value{footnote}}%
    \renewcommand*{\thefootnote}{\fnsymbol{footnote}}%
    \footnote[2]{#1}%
    \setcounter{footnote}{\value{daggerfootnote}}%
    \renewcommand*{\thefootnote}{\arabic{footnote}}%
    }

\begin{flushright}
\end{flushright}
\vspace{15mm}
\begin{center}
{\LARGE The Fuzzball Paradigm{\daggerfootnote{This manuscript is submitted as a chapter in the book: The Black Hole Information Paradox, A. Akil and C.
Bambi, editors.}}}

\vspace{10mm}

{\bf Samir D. Mathur$^{1}$ and Madhur Mehta$^2$
}

\blfootnote{$^{1}$ email: \href{mailto:mathur.16@osu.edu}{mathur.16@osu.edu}}
\blfootnote{$^{2}$ email: \href{mailto:mehta.493@osu.edu}{mehta.493@osu.edu}}

\vspace{2mm}

\b

Department of Physics

 The Ohio State University
 
Columbus,
OH 43210, USA

\b

\vspace{4mm}
\end{center}
\vspace{5mm}
\thispagestyle{empty}


%
%

\begin{abstract}
We describe the  puzzles that arise in the quantum theory of black holes, and explain how they are resolved in string theory.  We review how the Bekenstein entropy is obtained through the count of brane bound states. We describe the  fuzzball construction of black hole microstates. These states have no horizon and radiate from their surface like a normal body, so there is no information puzzle. We explain how the  semiclassical approximation is violated in gravitational collapse even though curvatures are low at the classical horizon. This violation happens because the collapse leads to a stretching of space that is {\it fast}: light does not have time to travel across the collapsing region to establish the `vecro' correlations needed in the quantum gravitational vacuum. These vecro correlations arise from the existence of virtual fuzzball fluctuations in the gravitational vacuum, and   are significant because of the large degeneracy of fuzzball states implied by the Bekenstein entropy.  We conjecture that  similar effects of fast expansion may be responsible for effects like dark energy and the Early Dark Energy postulated to explain the Hubble tension.
\end{abstract}

\section{The puzzles}

The quantum theory of black holes presents us with three interrelated  puzzles:

\b

(A) {\it The entropy puzzle:} Gedanken experiments suggest that black holes have an entropy \cite{bek}
\be
S_{bek}={c^3\over \hbar} {A\over 4G}\,.
\label{one}
\ee
But it has been argued that `black holes have no hair', so what is this entropy counting? At the other extreme, the bags-of-gold construction by Wheeler suggests that we can put an infinite amount of entropy inside the horizon. 

Thus, the puzzle is: {\it Does the entropy (\ref{one}) correspond to a count of black hole microstates through $S_{bek}=\log {\mathcal N}_{micro}$, in the way it would for a normal system described by statistical mechanics?}

\b

(B) {\it The information paradox:} In classical general relativity, matter collapses to the center of the black hole, leaving a vacuum around the horizon. Hawking \cite{hawking1,hawking2} found that in quantum theory, this vacuum region is unstable to the creation of entangled pairs. One member of the pair (which we will call $b$) escapes to infinity as `Hawking radiation'. The other member (which we will call $c$) has negative energy and falls into the hole, lowering its mass. The two members of the pair are in an entangled state, which we can schematically write as
\be
|\psi\rangle_{pair}={1\over \sqrt{2}}\left (|0\rangle_b|0\rangle_c+|1\rangle_b|1\rangle_c\right)\,.
\label{bits}
\ee
Thus, there is a monotonically growing entanglement between the emitted radiation and the remaining hole, leading to a sharp puzzle at the endpoint of evaporation. If the hole evaporates away, then we are left with radiation that is entangled, but there is nothing that this radiation is entangled {\it with}. Then this radiation cannot be described by {\it any} pure state in quantum theory, but only by a density matrix; thus we have a loss of quantum unitarity in the process of black hole formation and evaporation. This difficulty is known as the black hole information paradox. 

To evade this difficulty, one may postulate that quantum gravity effects prevent the further evaporation of the hole once the hole evaporates down to Planck size. The initial matter making the hole, and all the negative energy quanta $\{ c_i\}$ are then trapped in a tiny `remnant'. Since we could have started with a black hole of arbitrarily large size and evaporated down to Planck scale, the Planck mass remnant should be able to hold an arbitrarily large number of internal states. One model that has been proposed for remnants is in terms of a `baby universe' which is attached to the rest of spacetime  through a Planck sized neck. 

But remnants do not appear to be possible in string theory. For one, we believe we know all the states at the Planck scale; we just have a few strings and branes. For another, the conjecture of AdS/CFT duality rules out remnants. Suppose we have an object with energy $E_{grav}$ sitting at the center of an  $AdS_5$ space with curvature radius $R_{AdS}$. The dual theory is an $SU(N)$ gauge theory, with a finite value of $N$: we have $N\sim (R_{AdS}/l_s)^4 \sim (R_{AdS}/l_p)^4$ (we have assumed that the string coupling is a fixed number not scaling with $N$; thus in these estimates we are using $l_s\sim l_p$).  The gauge theory lives on a sphere $S^3$ whose radius can be taken to be any length $R_{CFT}$. Then the  remnant in the gravity theory with $E_{grav}\sim m_p$ is described in this gauge theory by a state of energy 
\be
E_{CFT}\sim {R_{AdS} E_{grav}\over R_{CFT}}\sim {N^{1\over 4}\over R_{CFT}}\,.
\ee
 The  gauge theory has finite $N$ and lives on a  finite volume sphere with radius $R_{CFT}$. In the finite  energy range $E\lesssim E_{CFT}$,  this gauge theory has only a finite number  of allowed  states. Thus, AdS/CFT duality does not allow the gravity theory to have an infinite number of states for a remnant with $m\lesssim m_p$. 

Given this, how do we resolve the problem of monotonically rising entanglement in string theory? Some people had initially hoped that the puzzle might be resolved through `small corrections'. Suppose there is a small correction of order $\epsilon$ to the state of each emitted pair; such small corrections to Hawking's computation can always arise from hitherto unknown quantum gravity effects. The number of emitted quanta is large: $N_{quanta}\sim (M/m_p)^2$, where $M$ is the black hole mass. Thus, if $\epsilon N_{quanta}\gtrsim 1$, one might hope that subtle correlations among this large number of emitted quanta might remove the monotonically growing entanglement that Hawking observed in his leading order calculations.

But the `Small corrections theorem' \cite{cern} shows that such a resolution is not possible. Suppose we assume 

(i) There is  small correction to the state of each pair $\{b_i, c_i\}$
\be
|\psi\rangle_{pair}={1\over \sqrt{2}}\left (|0\rangle_{b_i}|0\rangle_{c_i}+|1\rangle_{b_i}|1\rangle_{c_i}\right)+|\delta\psi_i\rangle, ~~~~\Big | |\delta \psi_i\rangle \Big |<\epsilon, ~~~~\epsilon\ll 1\,.
\label{utwo}
\ee

(ii) There is no significant change to the state of the  quanta $\{ b_i\}$ after they recede sufficiently far from the hole.

Then the entanglement $S_{ent}(N)$ of the radiation with the hole at step $N$ must keep growing monotonically as
\be
S_{ent}(N+1)>S_{ent}(N)+\log 2 -2\epsilon\,.
\label{uone}
\ee
In other words,  small corrections to the low energy dynamics at the horizon cannot resolve the difficulty; we need order {\it unity} corrections.

Thus, our puzzle is: {\it How should we resolve the puzzle of monotonically rising entanglement?}

\b

(C) {\it Breakdown of the semiclassical approximation:} Consider a shell of mass $M$ that is collapsing to form  a black hole.  Curvatures are low at the horizon scale
\be
{\mathcal R}\sim {1\over (GM)^2}\ll {1\over l_p^2}\,,
\ee
so one expects that usual low energy semiclassical dynamics will hold, and the shell will  pass smoothly through the horizon radius. If the shell goes through the horizon, how does its information ever come out? Points on the shell that are inside the horizon cannot send light signals to the horizon or exterior of the hole. Thus even though new physics can occur at the singularity, how can this new physics modify the pair creation at the horizon and resolve Hawking's paradox?

Thus, the puzzle is: \emph{What can lead to a breakdown of semiclassical dynamics at the horizon where curvatures are low?}

\b

In this article, we will describe how string theory resolves these puzzles. The resolution of puzzles (B) and (C) is obtained through the \textit{fuzzball paradigm}, which describes the structure and dynamics of black hole states in the theory. Before proceeding, we summarize the resolutions obtained in this paradigm.

In string theory, the Bekenstein  entropy does correspond to a count of microstates in the theory. Unlike the classical hole, which has all its mass at $r=0$, we find that these microstates are  horizon-size quantum objects called fuzzballs. The fuzzballs have no horizon and  radiate from their surface like a normal body; thus there is no analogue of the  monotonically rising entanglement (\ref{uone}).

But why does gravitational collapse lead to fuzzballs rather than the semiclassical hole? It is natural to expect that semiclassical dynamics will break down when curvatures exceed the Planck scale: ${\mathcal R}\gtrsim l_p^{-2}$. But as we noted above, this will not help us to resolve the information paradox; we need order unity corrections to dynamics at the {\it horizon}. We will find that there is indeed a second mode of breakdown of the semiclassical approximation, due to an effect that can be traced back to  the large value of the Bekenstein entropy (\ref{one}). 

Fuzzballs are bound states of the elementary excitations in string theory.   The existence of  a bound state in a field theory leaves an  imprint on the correlations in the {\it vacuum}. This effect is usually small in a normal quantum field theory, which has a limited set of bound states.  But  the correlations created by  fuzzballs are particularly significant, because (i) fuzzballs have an extended structure (with size of order horizon radius), and (ii) fuzzballs have a very high degeneracy (given by the Bekenstein entropy) which increases with the size of the fuzzball. These correlations created by virtual excitations of fuzzballs are termed `vecro correlations', for reasons we will see later. 

The existence of these vecro correlations in the gravitational vacuum leads to a breakdown of the semiclassical approximation in the process of gravitational collapse.  This process of  collapse is `fast', in the sense that light does not have time to travel across the collapsing region to establish the vecro correlations appropriate to the  vacuum. Thus, we do not reach the semiclassical vacuum solution inside the horizon region; instead we transition to a linear superposition of fuzzball states.

\section{Black holes in string theory}

The first difficulty we encounter in quantizing gravity is that loop amplitudes diverge, and cannot be renormalized. In string theory, on the other hand, we find that loop amplitudes are finite. This taming of  divergences can be traced back to the fact that string theory has an effectively infinite number of particle species: as we go to higher energies, we find new `particles' arising as new vibration modes of the string. 

The second issue we must deal with in any theory of quantum gravity is the diverging value  for the vacuum energy of quantum fields. In string theory, we assume supersymmetry of the underlying theory, so that at we have flat Minkowski space as a solution of the theory.

The most remarkable feature of string theory is that it is {\it unique}; i.e., it has no free parameters. The theory must live in 9+1 dimensions, and the fundamental objects --- strings and branes --- are uniquely determined along with their tensions and interactions. Interestingly, this unique and elaborate structure can be obtained, in retrospect, in the following way. The classical supergravity theory has certain discrete symmetries called T and S dualities. If we require that these symmetries continue to hold at the quantum level, we obtain string theory.

\subsection{Black hole entropy}\label{secentropy}

To make black holes in string theory, we must consider a bound state of the elementary objects in the theory. It is reasonable to think that the microstates of black holes involve strong quantum gravity effects, and that it would be hard to construct and count these microstates. But it turns out that in string theory, at least for extremal holes (holes with mass $M$ equaling the charge $Q$),  there is a way to count these states without understanding their explicit structure. The coupling in the theory is given by the value of a scalar field (the dilaton). Thus, we can study the bound states at weak coupling or at strong coupling. Extremal holes  are supersymmetric states of the theory, and an index which counts the states must remain the same as we change the coupling. It is relatively straightforward to count the states at weak coupling. This index argument then yields the count of states  at strong coupling, where the bound state is expected to yield a black hole.

Let us now turn to the bound states, which will give us our black hole. We compactify some spatial directions; it is convenient to study the theory in 4+1 noncompact directions to start with, and then extend the constructions to the case of 3+1 noncompact directions. 

Suppose we wrap a string around a compact circle. From the viewpoint of the dimensionally reduced theory, this string would appear as a point mass in the noncompact directions. The string also carries a charge, with strings winding one way around the circle having an opposite charge to strings winding the other way. Thus, in the dimensionally reduced theory, we obtain an object with mass $m_1$ and charge $q_1$.  Here $m_1=2\pi R T$, where $T$ is the tension of the string and $R$ is the radius of the compact circle. The subscript $1$ corresponds to the string being a 1-dimensional object, or `1-brane'; more specifically, this string is called an NS1 brane.   The string winding around the circle has $M=Q$ (mass equaling charge in appropriate units); this implies that the wound string  gives a supersymmetric state of the theory.

To make a black hole with a large mass, we must take a bound state of a large number $n_1$ of strings.   Such a bound state of NS1 branes has a simple description: we just have a string that winds $n_1$ times around the circle before closing on itself. Thus, we obtain an object with mass $n_1 m_1$ and charge $n_1 q_1$. 

It turns out however that such a mass source does not create a good black hole. The tension of the string causes the size of the compact circle to shrink to zero  at the location where the string is wound. In the dimensionally reduced theory, we find that the horizon area is zero: the horizon coincides with the singularity. Thus, the Bekenstein entropy is $S_{bek}={A\over 4G}=0$. This is at least consistent with the microscopics: the multi-wound string has just one state, so the microscopic entropy is $S_{micro}=\log 1 =0$.\footnote{Actually the spin of the string makes it a multiplet of $256$ states, but we will regard $S_{micro}=\log 256 \approx 0$ since this entropy does not scale with the number $n_1$ of strings in the bound state.}

To get a black hole we must add excitations which tend to {\it increase} the size of this compact circle. Consider a graviton moving around this circle, at a fixed location in the noncompact directions. This is a mode carrying momentum, so we call it a $P$ excitation. In the dimensionally reduced theory, such a Kaluza-Klein mode carries a mass $m_p={1\over R}$ and a charge $q_p=\pm {1\over R}$, where $R$ is the radius of the compact circle and the two possible signs of $q_p$ correspond to the two directions in which the graviton can orbit the circle. The bound state of  $n_p$ such gravitons is a single graviton carrying $n_p$ units of momentum, and in the dimensionally reduced theory corresponds to an extremal object with $M=Q={n_p\over R}$. We see that $M$ decreases as $R$ is increased, so the momentum modes tend to expand the compact circle.  

Thus, we try to make a bound state carrying both winding and momentum charges. The corresponding hole is called the 2-charge extremal hole, because it carries two types of charges -- one from winding strings and one from momentum modes. Suppose we assume that at strong coupling, the bound state of these charges forms a spherically symmetric solution of the supergravity equations, carrying the above mass and charges.  We seek to measure the horizon area of this solution and thus get $S_{bek}$ for the hole.

It turns out that if we use the leading order supergravity action, we still get a vanishing horizon area. But the effective action arising from string theory has higher derivative terms of the form $\alpha' {\mathcal R^2}$. Taking these into account, we find that there is indeed a small horizon (of radius $\alpha'^\h$), and we can compute  the Bekenstein-Wald entropy \cite{wald} of this horizon. Consider type IIB string theory with the compactification ${\mathcal M}_{9,1}\r {\mathcal M}_{3,1}\times K3\times T^2$ where $K3$ is a Calabi-Yau 4-manifold. One finds that \cite{dabholkar}
\be
S_{bek-wald}=4\pi \sqrt{n_1\,n_p}\,.
\ee
We now wish to see if this gravity result can be reproduced from the microscopic description of the bound state. At weak coupling, it is easy to understand our bound state. We have already seen that the bound state of $n_1$ strings is a single string with winding $n_1$; this string has a total length $L_T=2\pi R\, n_1$. The momentum-carrying modes bound to this string  take the form of  traveling waves on the string. The waves on a string of length $L_T$ carry momentum in units of $2\pi/L_T$. Thus, we write the total momentum on the string as
\be
P={n_p\over R}={2\pi\, n_1\,n_p\over L_T}\,.
\ee
A quantum in the $k\,$th harmonic carries a momentum $p={2\pi\, k\over L_T}$. If we have $N_k$ units of the excitation in harmonic $k$, then we will get
\be
\sum_{k=1}^\infty k N_k = n_1n_p\,.
\label{bfiveq}
\ee
Note that each different direction of vibration counts as a different `flavor' of excitation. The different ways of partitioning the total momentum $P$ among these excitations gives the microscopic number of states ${\mathcal N}_{micro}$, and we obtain the microscopic entropy as $S_{micro}=\log {\mathcal N}_{micro}$. We find, for the compactification mentioned above \cite{sen1,sen2,dabholkar},
\be
S_{micro}=4\pi \sqrt{n_1n_p}=S_{bek-wald}\,.
\label{bfive}
\ee

The 2-charge extremal hole is also called the `small black hole' because the horizon radius is $\sim \alpha'^\h$, a microscopic length in the theory. To get a larger horizon, we can add a third kind of charge. We take the compactification ${\mathcal M}_{9,1}\r {\mathcal M}_{4,1}\times {\mathcal M}_4\times S^1$, where ${\mathcal M}_4$ is a 4-manifold, which can be K3 or $T^4$. We wrap strings and momentum modes around the $S^1$ the same way we did for the 2-charge hole, but now also add $n_5$ NS5-branes wrapping ${\mathcal M}_4 \times S^1$. The corresponding classical hole is a traditional Reissner-Nordstrom hole, whose Bekenstein entropy can be found by using the leading order action ${1\over 16\pi G} \int dx \sqrt{-g} {\mathcal R}$; the higher order Wald corrections are subleading. We find
\be
S_{bek}={A\over 4G}=2\pi\sqrt{n_1n_5n_p}\,.
\ee
The microscopic entropy can be again computed at weak coupling, and one finds \cite{sv}
\be
S_{micro}=2\pi\sqrt{n_1n_5n_p}=S_{bek}\,.
\ee

Extremal holes do not radiate. To understand radiation from the hole and the resolution of the information paradox, we should look at non-extremal holes. We can make a non-extremal microstate by taking momentum modes $P$ carrying momentum $n_p$  in one direction and also momentum modes $\bar P$ carrying momentum $\bar n_p$ in the opposite direction. The microstate will no longer be supersymmetric, so we cannot rigorously argue that the count of states at weak coupling must agree with the count at strong coupling. We can avoid violating supersymmetry `too much' by taking $n_p, \bar n_p\ll n_1, n_5$. Then we find a microscopic entropy at weak coupling
\be
S_{micro}=2\pi\sqrt{n_1n_5}\left ( \sqrt{n_p}+\sqrt {\bar n_p}\right )\,.
\label{uttwo}
\ee
We compare this microscopic entropy to $S_{bek}={A\over 4G}$, where $A$ is computed using the horizon area of a black hole carrying the mass and charges of the microstates; note that the momentum charge carried by the hole is $n_p-\bar n_p$. We again find agreement: $S_{micro}=S_{bek}$ \cite{callanmalda}.

Remarkably, a natural extension of the weak coupling expression (\ref{uttwo}) continues to give an exact agreement with the Bekenstein entropy even for  holes which are arbitrarily far from extremality \cite{hms}.  There is no clear reason known why this agreement should hold.  Nevertheless, in view of all these agreements, it is reasonable to say that in string theory, the Bekenstein-Wald entropy indeed represents a count of microstates of the hole, in exactly the way we understand entropy in statistical physics as a count of microstates.

\subsection{Radiation from microstates}\label{secrad}

Extremal holes cannot radiate, but near-extremal holes radiate, and we can use them to approach the Hawking puzzle. At strong coupling where we expect black holes, we can perform a computation similar to the one that Hawking did, and obtain a radiation rate $\Gamma_H(\omega)$ which depends on the energy $\omega$ of the radiated quantum. On the microscopic side,   we can look at the weak coupling description which gave the entropy (\ref{uttwo}). In this description we have massless momentum carrying modes $P$ and $\bar P$ running in opposite directions along the $S^1$ on which the branes  are wrapped. These modes can collide, cancelling their momentum charge, and leading to the emission of a graviton from the string bound state. The interaction vertex $\hat V$ for this process  thus gives an operator of the form
\be
V(\omega) \hat a_P(p)\hat a_{\bar P}(p)\hat a^\dagger_h(\omega)\,,
\label{boneq}
\ee
where $\hat a_P(p)$ annihilates a $P$ excitation with momentum $p$ and energy $p$, $\hat a_{\bar P}(p)$ annihilates a $\bar P$ mode carrying momentum $-p$ and energy $p$, and $\hat a^\dagger_h(\omega)$ creates a graviton with no momentum along the $S^1$ and energy $\omega=p+p=2p$. The radiation rate $\Gamma$ is then computed by multiplying $V(\omega)$ by the occupation numbers $N_p$ and $\bar N_p$ of the relevant $P$ and $\bar P$ modes. Schematically:
\be
\Gamma_{micro}(\omega) = V(\omega) N_p({\omega\over 2}) \bar N_p({\omega\over 2})\,.
\label{bone}
\ee
The entropy (\ref{uttwo}) reproduced the Bekenstein entropy when we summed over all the allowed configurations of the $P, \bar P$ modes. By the usual arguments of statistical mechanics, we see that the entropy (\ref{uttwo}) will be reproduced by using  thermal distributions for $N_p, \bar N_p$, with temperatures corresponding to the total energies in the $P, \bar P$ sectors. Substituting these  thermal distributions for $N_p, \bar N_p$ in the radiation rate (\ref{bone}), we find \cite{dasmathur, maldastrom}
\be
\Gamma_{micro}(\omega)=\Gamma_H(\omega)\,,
\label{btwo}
\ee
where $\Gamma_H(\omega)$ is the radiation rate obtained from the semiclassical Hawking computation. This agreement (\ref{btwo}) suggests that string theory is on the correct track to understand black hole radiation. But note that the microscopic emission (\ref{bone})  arises from a unitary process, with excitations on the brane colliding and emerging as gravitons. By contrast, $\Gamma_H$ is obtained from the usual process of pair creation  from the vacuum, which leads to the information paradox. This suggests there is something wrong with the strong coupling picture of the black hole,  and we will now see that such is indeed the case.

\section{Constructing the microstates --- fuzzballs}

In the above discussion we counted black hole microstates by an indirect argument --- using supersymmetry to relate the state count at weak coupling to the count  at strong coupling. At weak coupling we could count states, but the states did not make black holes. At  strong coupling we just used the standard metric of the hole and computed the value of Bekenstein's area entropy. But this black hole metric has a horizon, and horizons lead to the information paradox.  To understand how the paradox is to be resolved, we must construct the microstates at {\it strong} coupling, and examine their structure. This is what we will proceed to do now. We will start with the simplest black hole: the 2-charge extremal hole with entropy (\ref{bfive}). The expressions for entropy were similar for all holes, so we expect that any lessons that we extract from the simple 2-charge extremal  case would hold, at least qualitatively, for all holes.

Recall that the 2-charge extremal hole was made by winding a string around a compact circle, and adding momentum along this string. All the states of the hole are given by the different traveling waves on this string. Thus, we need to construct the gravity solution created by   strings carrying  traveling waves.

With a view to study the 3-charge hole later, we compactify spacetime as ${\mathcal M}_{4,1}\times S^1\times T^4$. The $S^1$  is parametrized by  $0\le y<2\pi R$.  The $T^4$ has volume $(2\pi)^4 V$ and is parametrized by $z_a, a=1, \dots 4$. 

 First, consider a static string source, with the string wound  along the $S^1$.   We assume that this string source is made up of  $n_1$ fundamental strings, all at the same position. The classical solution for such a string source is known,
 \begin{align}
ds^2_{string}&=H(-dudv)+\sum_{i=1}^4 dx_idx_i+\sum_{a=1}^4 dz_adz_a\,, \nonumber
\\
B_{uv}&=-{1\over 2}(H-1)\,, \quad
e^{2\phi}=H\,,\quad
H^{-1}=1+{Q_1\over r^2}\,,\quad
Q_1= \,\frac{g^2 \alpha'^3 }{V}n_1.\label{f1}
\end{align}
Let us now add a traveling wave to this string. There are two important aspects to note. First, the fundamental string of string theory has no longitudinal vibration modes; it can have only transverse vibrations. Thus, when we add momentum charge $P$ in the form of a traveling wave, the string necessarily bends away from its central position, spreading over some transverse spatial region. In other words, the string  does not remain a point source in these  transverse directions. Second, the string is multiwound, with winding $n_1$. Thus, the transverse displacement of the string has to return to its initial value, not after $\Delta y=2\pi R$ but after $\Delta y = 2\pi R\, n_1$. Under such a wave, the strands separate from each other in general, and we find $n_1$ string sources, each carrying a wave traveling in the $y$ direction at the speed of light. The metric for such a set of string sources is known:
 \begin{align}
ds^2_{string}&=H[-du\,dv+Kdv^2+2\,A_i\, dx_i\, dv]+\sum_{i=1}^4 dx_i\,dx_i+\sum_{a=1}^4 dz_a\,dz_a\,,\nonumber
\\
B_{uv}&=-{1\over 2}[H-1], ~~\quad B_{vi}=HA_i\,,\quad
e^{2\phi}=H\,,\nonumber
\\
H^{-1}(\vec x, y,t)&=1+\sum_s{Q_1^{(s)}\over |\vec x-\vec F^{(s)}(t-y)|^2}\,,\quad K(\vec x, y,t)=\sum_s{Q_1^{(s)}|\dot{\vec F}^{(s)}(t-y)|^2\over |\vec x-\vec F^{(s)}(t-y)|^2}\,,\nonumber
\\
A_i(\vec x,y,t)&=-\sum_s{Q_1^{(s)}\dot F^{(s)}_i(t-y)\over |\vec x-\vec F^{(s)}(t-y)|^2}\,.
\label{fpmultiple}
\end{align}
In studying black holes, we are interested in the limit $n_1, n_p\gg 1$. In this limit we may replace the sum in (\ref{fpmultiple}) by an integral
\be
\sum_{s=1}^{n_1} \r \int _{s=0}^{n_1} ds = \int_{y=0}^{2\pi R\, n_1}{ds\over dy} dy
\ee
Since the length of the compact $S^1$  is $2\pi R$ we have
${ds\over dy}={1\over 2\pi R}$. 
Also, since the vibration profile is a function of $v=t-y$ we can replace the integral over $y$ by an integral over $v$. Thus, we have
\be
\sum_{s=1}^{n_1}\r {1\over 2\pi R}\int_{v=0}^{L_T }dv\,,
\ee 
where 
$L_T=2\pi R n_1$
is the total range of the $y$ coordinate on the multiwound string. Finally, note that
$Q_1^{(s)}={Q_1\over n_1}$. With all this, the solution (\ref{fpmultiple}) becomes
\begin{align}
ds^2_{string}&=H[-du\,dv+K\,dv^2+2\,A_i\, dx_i\, dv]+\sum_{i=1}^4 dx_i\,dx_i+\sum_{a=1}^4 dz_a\,dz_a\,,\nonumber
\\
B_{uv}&=-{1\over 2}[H-1], \quad B_{vi}=HA_i\,,\quad e^{2\phi}=H\,,\nonumber
\\
H^{-1}&=1+{Q_1\over L_T}\int_0^{L_T}\! {dv\over |\vec x-\vec F(v)|^2}\,,\nonumber\\
K&={Q_1\over
L_T}\int_0^{L_T}\! {dv (\dot
F(v))^2\over |\vec x-\vec F(v)|^2}\,,\nonumber\\
A_i&=-{Q_1\over L_T}\int_0^{L_T}\! {dv\dot F_i(v)\over |\vec x-\vec F(v)|^2}\,.
\label{ttsix}
\end{align}

In fig.\ref{fig:fuzzballstrings} we give a graphic description of the steps above. If we do not take into account the transverse displacements of the string, we get the `naive' geometry describing a black hole with horizon (after taking account $\alpha'{\mathcal R^2}$ corrections). Taking into account the actual transverse displacements gives the fuzzball solutions, which have a different geometry for each  allowed distribution of momentum on the string.

The solutions (\ref{ttsix}) give the microstates of the 2-charge extremal hole made with string winding charge (usually called NS1) and momentum charge (usually called P) \cite{lm4}. These are all the microstates where the string is allowed to oscillate in the 4 noncompact directions $\vec x$. The solutions with vibrations in the compact directions $z_a$ are similar, and were given in \cite{lmm}, and the set of solutions for the heterotic string were given in \cite{skenderis}. Special cases of these metrics were noted earlier. Solutions with $\vec F(v)$ lying along a circle were found in \cite{bdkr, mm, multiwound, martinec}. Solutions with excitations lying in a plane (supertubes) were found in \cite{supertubes}. 

Let us now analyze the nature of these solutions, and the lessons that we can draw from them about the structure of black holes.

\section{Lessons from the fuzzball construction}
\begin{figure}
    \centering
    \includegraphics[width=0.8\linewidth]{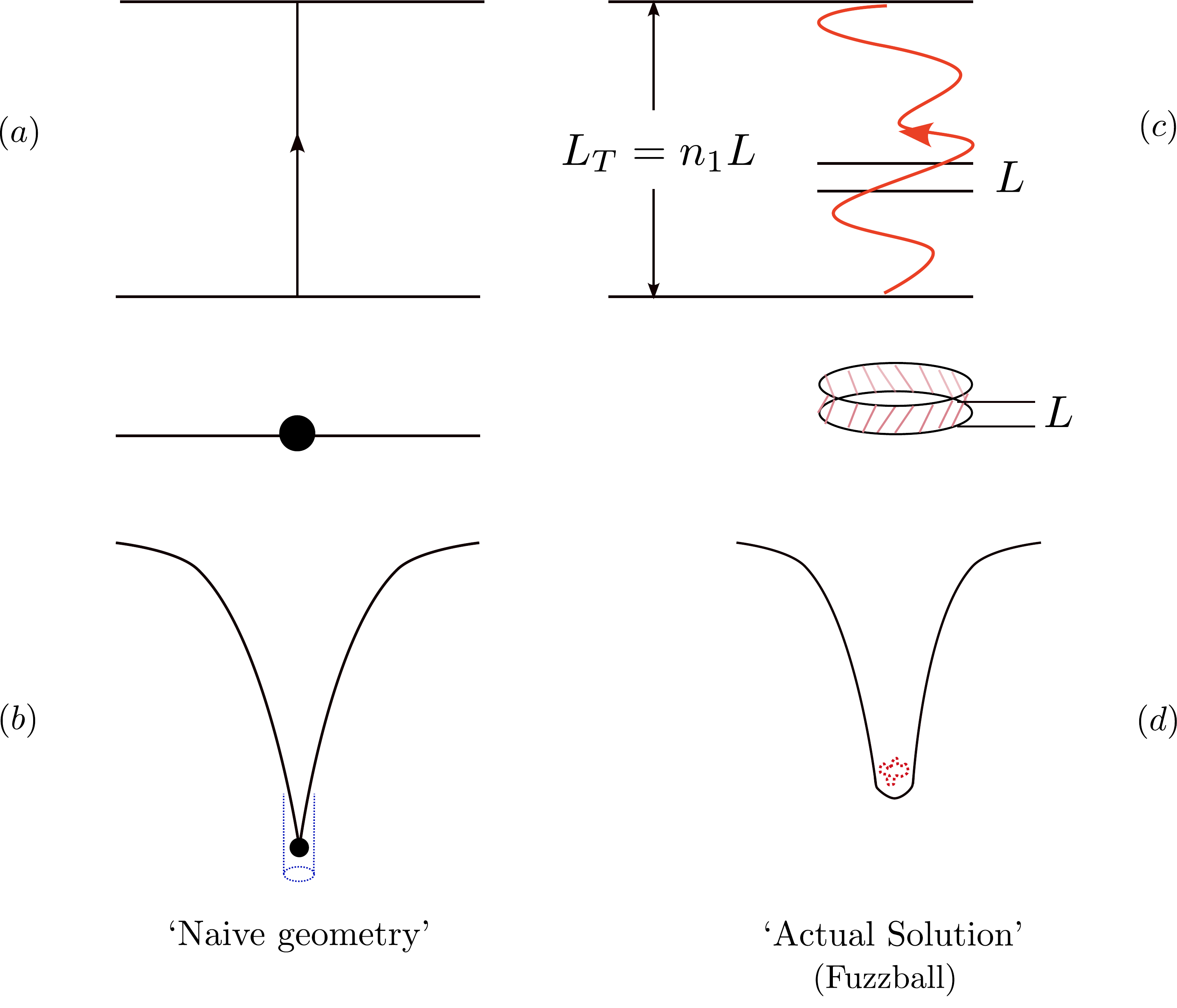}
    \caption{(a)  If a string could carry momentum as a longitudinal wave, then it would generate a spherically symmetric geometry. (b) This `naive' geometry would have a singular horizon at leading order, but $\alpha'$ corrections turn this singularity into a regular horizon. (c) The actual string of string theory has only transverse vibrations, which break spherical symmetry and cause the string to spread out over these transverse directions. (d) The resulting geometry is a `fuzzball' with no horizon.}
    \label{fig:fuzzballstrings}
\end{figure}
Analysis of fuzzball construction above suggests the following nature for black hole microstates in string theory:

\b

(A) {\bf Absence of a horizon:} A little analysis shows that none of the microstates (\ref{ttsix}) have a horizon. Along the curve $\vec x=\vec F(v)$, the redshift goes to infinity, but there is no closed trapped surface, and light rays can escape to infinity from any point in these geometries.

\b

 (B) {\bf Lack of spherical symmetry:} Note that the solutions (\ref{ttsix}) are not spherically symmetric. If we {\it impose} spherical symmetry on the low energy gravity solution, then we get the `naive' metric pictured in fig.\ref{fig:fuzzballstrings}. This naive metric is not realized in string theory, because it is not possible to add a momentum carrying wave to the string without breaking spherical symmetry. This in turn follows from the fact that the fundamental string has no longitudinal vibration mode; it only allows transverse vibrations, and the polarization of this vibration at any point along the string must break the spherical symmetry in the angular $S^3$ directions or the symmetry in the $T^4$. 

\b

(C) {\bf The relation between entropy and area: } Given that there is no horizon, one might wonder what happens to the Bekenstein entropy relation (\ref{one}). From fig.\ref{fig:fuzzballstrings} we see that the microstates have the structure of flat space at infinity, then a `neck' which leads to a `throat', and this throat ends in a `cap'. The geometries are almost identical everywhere except in the cap; thus this cap contains the detailed information of the choice of microstate. Let us draw a surface that surrounds the region where the typical microstates differ significantly from each other; this surface can be said to enclose the `information' of the microstate. We find that the area of this surface satisfies \cite{lm5}
\be
{A\over G}\sim \sqrt{n_1n_p}\sim S_{micro}\,.
\label{utone}
\ee
We can also consider the subset of the microstates   which have a large angular momentum $J$, with $J\gtrsim (n_1n_p)^{1\over 4}$. The entropy for such microstates can be computed in a manner similar to (\ref{bfiveq}), and we find
\be
S_{micro} = 2\pi\sqrt{2} \sqrt{n_1n_p-J}\,.
\ee
In this case, the geometries differ from each other over a tube shaped region, so we get the analogue of a black ring. The surface area of this tube is  found to satisfy \cite{lm5}
\be
{A\over G}\sim\sqrt{n_1n_p-J}\sim S_{micro}\,.
\ee
Thus, it appears that the Bekenstein entropy relation tells us the number of orthogonal gravity solutions that we can fit in a given region. We can investigate this further by restricting attention to only that subset of microstates which `fit' in a small ball of surface area $A'=\mu A$ where $\mu<1$ and $A$ is the bounding area of the generic microstate,  appearing in (\ref{utone}).  In other words, we look at states that differ significantly from each other only inside the ball of area $A'$; these are a special subset of all microstates. The count of these states turns out to give \cite{phase}
\be
S_{micro}\sim \mu \sqrt{n_1n_p}\sim {A'\over G}\,,
\ee
so we again find a Bekenstein-type relation.

Note that the geometries depend on several continuous parameters $R, V$ and $ g$. In the ratio $A/G$ in (\ref{utone}) all these variables cancel out, yielding the entropy which depends only on the charges. It is not clear why exactly we have this cancellation, so it is not easy to answer the question of why the surface area is related in the above way to entropy. But these computations do indicate that the gravity solutions we have constructed are packed to maximal density; i.e., we cannot fit more than $Exp[S_{bek}]$ orthogonal gravity states in a region that is bounded by a surface smaller than that given by the relation $A/4G=S_{micro}$.

The  computations of $S_{micro}$  in section\,\ref{secentropy} were performed at weak coupling, while the fuzzball solutions (\ref{ttsix}) were constructed at strong coupling. In \cite{rychkov}, the space of fuzzball solutions (i.e., the solutions obtained at strong coupling) was quantized and shown to yield the entropy $S_{micro}$.

\b

(D) {\bf 3-charge holes:} We have obtained all the microstates of the 2-charge extremal hole, and found that they are all fuzzballs. We now wish to consider more complicated holes like the 3-charge hole. This hole can be made by adding a 5-brane charge to the winding and momentum charges that we took for the 2-charge hole.

Before looking at the 3-charge hole, it is useful to take the 2-charge fuzzball solutions (\ref{ttsix}) and apply some  discrete symmetry maps --- called S and T dualities --- to these solutions. After the dualities, the charges will change as follows
\be
{\rm NS1\,  charge }~\r ~ {\rm NS5\, brane\,  charge\, } , ~~~{\rm  P\,  charge}~~\r~~{\rm  NS1\,  charge}\,.
\ee
  Thus, we can map the fuzzballs (\ref{ttsix}) with NS1-P charges of the fuzzballs to NS5-NS1 charges.\footnote{Most of the black holes constructions discussed below were done using D5 and D1 branes, but here for simplicity we apply an S-duality map to those constructions and work with the NS5-NS1 bound state.}    Under this map we find that the singularity in (\ref{ttsix}) along the locus $\vec x = \vec F(v)$ gets removed, and we obtain metrics that are smooth everywhere. The points along the curve $\vec x = \vec F(v)$ become centers of Kaluza-Klein monopoles, which have a nontrivial topology but no singularity anywhere. More generally, microstates can have string and brane sources, as well as nontrivial topological features. They will in general be described by quantum wavefunctionals, and only a subset of  fuzzballs can be well-approximated by a classical metric. The important point however is that no microstate has the traditional  horizon with the semiclassical vacuum state in its vicinity. 

In \cite{mss, lunin, gms1,gms2}, simple examples of extremal solutions were constructed with NS1, NS5 and P charges; i.e. these solutions carried the charges of the Strominger-Vafa black hole.  In each case, the solution was a fuzzball; i.e, there was no horizon. Amplitudes computed for particular states in the dual CFT agree with amplitudes computed in the corresponding gravity microstates \cite{russo}. 

In the solutions of \cite{gms2}, four of the directions were compactified to a torus $T^4$, which was trivially tensored with the other directions. Dimensionally reducing on this $T^4$ gives a supersymmetric solution in 5+1 dimensions, where one of the directions is a $S^1$ parametrized by $0<y<2\pi R$, one direction is time $t$, and four space directions are noncompact. There was a general result \cite{gauntlett} that supersymmetric solutions in 5+1 dimensions could be written as a 4-dimensional hyperkahler base times a 1+1 dimensional fiber along the directions $t,y$. In \cite{giustomathur}, the solutions of \cite{gms2} were decomposed in this base$\times$fiber fashion, and a surprise was found. Hyperkahler manifolds have signature $++++$, but in this decomposition the hyperkahler base had signature $++++$ in one region (connected to spatial infinity) and signature $----$ in an interior region. The base$\times$fiber split degenerated at the boundary of these two regions, but the overall 6-dimensional geometry remained smooth everywhere, with signature 5+1. Thus, we again have an example of a fuzzball geometry where the compact directions are not trivially fibered with the noncompact directions, though the overall manifold remains regular.

\b

(E) {\bf Finding large sets of fuzzball solutions:} In \cite{bw01}, a general method was found to find extremal solutions of supergravity with this kind of base$\times$fiber split. This work led to an intense program which was very successful in constructing a large number of families of microstates for the 3-charge extremal hole \cite{berglund, bw02, bwreview}. These constructions describe  microstates of black holes, black rings, and combinations of black holes and black rings. Some solutions have brane sources in the form of supertubes in the `cap' region. The arbitrary 1-dimensional curve $\vec F(v)$ appearing in the 2-charge extremal solutions (\ref{ttsix}) has been extended to an arbitrary  2-dimensional surface in 3-charge extremal solutions \cite{bw1,bw2}. An important aspect of these constructions was that they  succeeded in constructing  microstates in the `black hole regime': while earlier 3-charge fuzzball constructions were typically solutions that were `overrotating', these solutions have the macroscopic parameters required to describe a nonrotating 3-charge hole. The large number of microstates found in this program went a long way towards establishing the non-uniqueness of the standard black hole geometry in theories with extra dimensions and,  in particular, in string theory.

\b

(F) {\bf Non-extremal fuzzballs and Hawking radiation:} Extremal holes do not radiate. Thus while extremal fuzzballs tell us something about the structure of black holes, we should construct non-extremal microstates to understand radiation from the hole and the resolution of the information paradox.  

In \cite{jmart} a family of non-extremal D1D5P microstates were constructed. These solutions were again found to be fuzzballs; i.e., they had no horizon. They did have an ergoregion, and the emission from this ergoregion was computed in \cite{myers}. In \cite{cm1,cm2, virmani} it was found that this emission spectrum was exactly the spectrum of Hawking radiation that was expected from these very non-generic microstates. To understand the nature of this agreement, we proceed in the following steps:

\b

(i) In the CFT, we have an emission vertex $  \hat V$ (eq.\,(\ref{boneq})) which causes a transition from excitations on the weakly coupled NS1-NS5 bound state to gravitons that escape this bound state. If we have a generic thermal distribution of excitations on the bound state, then we multiply this vertex by Bose/Fermi distributions as in (\ref{bone}) to obtain the emission rate  from the generic brane bound state at weak coupling.  As noted in section \ref{secrad}, this emission spectrum  exactly matches the spectrum of Hawking radiation from the corresponding black hole. 

\b

(ii) With the fuzzball constructions,  we have explicit gravity microstates (without horizon) for specific cases, together with knowledge of the corresponding weak coupling CFT state. Thus,  we can compute  the CFT emission from a specific microstate by multiplying the vertex $V$ with the occupation numbers for excitations on that {\it particular} microstate.  What we find is that this emission rate from the particular microstate at weak coupling exactly matches the emission from the corresponding gravity solution (i.e. fuzzball) obtained at strong coupling. 

\b

(iii) In short, the same weakly coupled CFT computation which reproduces the Hawking rate from a black hole, also reproduces the emission from specific fuzzballs.  But while Hawking's emission led to a loss of information, the emission from the fuzzball is completely unitary, and we can see the details of the specific fuzzball imprinted on the emission spectrum.  Of course, since these fuzzballs are very nongeneric states, the emission profile in these cases is not at all like thermal radiation.  But these computations do suggest that when we are able to compute radiation from generic fuzzballs, we will find that the radiation is unitary but still with the emission spectrum found by Hawking.

\b

Other  classes of nonextremal microstates have been constructed in  \cite{rotating, bolts, nonex}.

\b

(G) {\bf The general nature of fuzzball solutions:} We end this section with some general comments on the nature of fuzzballs. The Buchdahl theorem says that if a ball of perfect fluid has a radius $R<9/4 GM$, then it must collapse into a black hole. On the other hand, we expect fuzzballs to have a surface that is just order Planck length outside the horizon radius \cite{ghm}. What novel effects in string theory would allow such solutions?

A black hole in 3+1 dimensions will have 6 compact directions in string theory.  We have already seen above that in fuzzballs, the noncompact directions are not trivially tensored with compact directions. As a toy example, consider  3+1 noncompact spacetime with one  additional compact circle.  Then the vacuum Einstein equations have a solution given by  Euclidean Schwarzschild times a trivial  a time direction
 \be
 ds^2=-dt^2+(1-{r_0\over r})dy^2 + {dr^2\over 1-{r_0\over r}}+ r^2(d\theta^2+\sin^2\theta d\phi^2)
 \ee
 where $r\ge r_0$ and the radius of the compact circle $y$ is $4\pi r_0$.  If we dimensionally reduce on the $y$ circle, we get 3+1 spacetime coupled to a scalar field. This scalar field has the usual sign of the stress tensor, and is nonzero everywhere, diverging as $r\r r_0^+$.  From a 3+1 dimensional perspective, we can ask:  why is this distribution of energy density not collapsing inwards under its own gravity?  
 
 If we take a different radial function for the same compactification, we can obtain something even more startling:  Witten's bubble of nothing \cite{bubble}.  This is also a vacuum solution of the Einstein's equations in 4+1 dimensions, and from a 3+1 dimensional perspective the matter is a scalar field with the usual stress tensor. Yet instead of collapsing inwards under gravity, the solution is exploding outwards. 
 
 As noted in \cite{what}, in each of the above cases a key point is that the manifold is not a trivial product of 3+1 dimensional space and a circle: the circle degenerates on a $S^2$, creating a cigar shaped `cap' at that location. In this situation the stress tensor of the scalar field is highly non-isotropic, with positive pressure in the radial direction and negative pressures in the angular directions; thus the pressure is very different from the isotropic one assumed in Buchdahl's theorem. These toy examples show how the usual intuition of gravity can be violated when we have compact dimensions.\footnote{These toy models cannot be directly realized in string theory because the cigar geometry is not compatible with the presence of fermions in the theory. A fermion that is periodic around the compact circle at infinity will have a singularity at the tip of the cigar.} 
 
 As noted in (F) above,  fuzzball solutions in appropriate duality frames have compact directions that are not trivially tensored with noncompact directions. In  other duality frames this topological nontriviality shows up as a string or brane source: for example in the NS1-P frame the 2-charge extremal fuzzballs had a string source, while in the NS1-NS5 duality frame this source was replaced by a smooth KK-monopole tube. Thus, it is the special features of string theory which allow fuzzballs to exist. 
 
 In \cite{gibbonswarner} it was shown how these features bypass the usual no hair arguments for black holes. In \cite{heidmann} solitonic stars were constructed to model neutral black holes, with compact directions pinching off in the cigar shaped manner mentioned above. 
 
 Far outside from the fuzzball, we expect to get the usual metric of the black hole. What happens as we approach the horizon radius? To get an idea of this, let us examine the nature of the extremal fuzzballs which have been constructed.  At different locations in the fuzzball, we have a string theory source, which is always a $\h$-BPS object. The overall configuration is ${1\over 4}$-BPS or ${1\over 8}$-BPS, depending on the hole whose microstates we are constructing \cite{bala}. Some part of these source charges cancel out in the overall charge of the fuzzball; we call these dipole charges. The other parts add up to yield the net charge of the fuzzball. As we approach one of these source charge locations, the geometry becomes very asymmetric: the metric along the brane source shrinks to zero by the tension of the brane, and the metric perpendicular to the brane source expands to infinity. 
 
 A similar situation is noted with metrics that have been proposed for some neutral Extremely Compact Objects (ECO) \cite{ecometric}.  In the traditional black hole metric, as we approach the horizon in a radial direction, we find Rindler space. In a suitable frame around any point near the horizon, the time direction and the radial direction make up the right Rindler wedge, while the directions along the horizon separate out with a metric of the form  $ds_\perp^2\approx dX_1^2+\dots dX_{d-1}^2$. But in the ECO geometry in \cite{ecometric}, the metric in some of these horizon directions diverges  as we approach the horizon radius, while in other directions it shrinks to zero. This behavior is also observed for the solitonic stars mentioned above. We can regard these local solutions as similar to the vacuum Kasner metrics for cosmology. In the Kasner solutions, the metric has the form 
 \be
 ds^2=-dt^2+\sum_i t^{2\alpha_i} dX_i^2
 \ee
 To get the possible geometries near the fuzzball surface, we  replace the $t$ coordinate in the Kasner solutions by the distance from the horizon $s$.  The metric then takes the form
 \be
 ds^2 \approx  -s^{2\alpha_0} dt^2 + \sum_i s^{2\alpha_i} dX_i^2
 \label{ffone}
 \ee
 We again find the conditions on the exponents that are found for the Kasner case:
 \be
 \alpha_0+\sum_i\alpha_i=1, ~~~~\alpha_0^2+\sum_i \alpha_i^2=1
 \label{bkl}
 \ee
 (The metric in \cite{ecometric} for example satisfies these relations.)
 
 For the case of cosmology, the BKL estimate \cite{bkl} shows that as $t\r 0$, we get a random orientation of the spatial axes determining the directions $X_i$.  That is, these directions and their corresponding exponents change very rapidly from position to position. If a similar situation holds for the metric (\ref{ffone}), then we will get a large family of allowed solutions as we approach the horizon, and this can be a manifestation of the large entropy of the hole. When compact directions $X_i$ in (\ref{bkl}) have positive exponents, they shrink towards zero size near the horizon. Branes wrapping these directions can become very light, and the fuzzball surface will become a region with large quantum fluctuations.

\section{Recovering black hole thermodynamics}

Let us summarize what we have seen above. The presence of a horizon
leads to the information paradox. The small corrections theorem
makes this into a precise conflict: if semiclassical physics around a horizon
emerges, even as a leading order approximation, then black hole
evaporation must end in either a loss of unitarity or the
formation of remnants with unbounded degeneracy. String theory
escapes this problem by not allowing the formation of a horizon: if
we make a bound state of the fundamental objects in the theory ---
strings and branes --- then this bound state is a horizon sized quantum
object --- a fuzzball --- with no horizon. The fuzzball radiates from
its surface like a normal body. Thus,
the black hole is just like a `string star', and there is no information paradox.

But now we are faced with a  different question. The
traditional semiclassical black hole has an elegant thermodynamics. For example, in
3+1 dimensions, the Schwarzschild  hole has a temperature
\be
T_H={1\over 8\pi GM}\,,
\label{yone}
\ee
and an entropy
\be
S_{bek}={A\over 4G}\,.
\label{ytwo}
\ee
 Further, this semiclassical hole
radiates like a black body: i.e., the radiation rate $\Gamma_{BH}(l,m,\omega)$ for a partial wave $Y_{l,m}$ of a massless scalar at  
frequency $\omega$ is related to the absorption probability ${\cal P}(l,m,\omega)$ by the  relation of detailed balance:
\be
\Gamma_{BH}(l,m,\omega)d\omega= {{\cal P}(l,m,\omega)\over e^{\omega\over T_H}-1} {d\omega\over 2\pi}\,.
\label{ythree}
\ee
The absorption probability ${\cal P}(l,m,\omega)$ depends only on the geometry of the hole outside the horizon.

But if the black hole is just a string star with no horizon, then
why should it have the properties (\ref{yone})-(\ref{ythree})? Hawking's derivation of the temperature (\ref{yone}) relied on his picture of pair creation at the horizon. If we have no horizon, then we resolve the information paradox.  But in the process, do we lose the elegance of black hole  thermodynamics?

As we will now note, if an object is sufficiently compact (i.e., it
has a radius that is sufficiently close to the horizon radius $R=2GM$), then it
{\it must} have the temperature, entropy, and radiation rate given by black
hole thermodynamics \cite{eco1, eco2}. We use the term Extremely Compact Object (ECO) to describe an object with a surface which is a very small distance $s$ outside the horizon radius $r=2GM$. ECOs have been of interest in general theories of quantum gravity -- not just string theory -- and the argument below will be a general one which applies to any ECO.

Consider a spherically symmetric ECO with mass $M$ as measured from
infinity. To get the essence of the argument, consider an extreme
case: the surface of the ECO is at a proper distance $s_{ECO}\sim l_p$
outside the horizon radius $R=2GM$, and the temperature is $T_{ECO}=0$.
We assume nothing about the dynamics inside the ECO surface, but
outside this surface we are allowed to use the semiclassical dynamics
of quantum fields on curved space. It is possible to argue that this
semiclassical region will have a stress tensor that equals, to leading
order, the stress tensor of the Boulware vacuum. This vacuum has a
vacuum energy density $\rho_v\sim -{1\over s^4}$, at a distance $s$
from the horizon radius. Integrating $\rho_v$ from the surface of the
ECO to infinity, we find that this vacuum energy contributes a mass
$M_{v}\sim -\alpha M$, with $\alpha$ a positive constant of order
unity.

Since the mass at infinity is $M$, the mass inside the ECO surface is
$(1+\alpha)M$, which corresponds to a horizon radius $2(1+\alpha) GM$.
Thus, the semiclassical region just outside the ECO is deep inside this
horizon radius. The geometry in this semiclassical cannot be time independent, due to the `inward pointing' behavior of light cones inside a horizon. Thus, we
conclude that an ECO with surface at $s_{ECO}\sim l_p$ and $T=0$ cannot
exist.

A similar argument works for other values of $T_{ECO}$, except for
$T_{ECO}= T_H$, where $T_H$ is the Hawking temperature. When $T=T_H$,
the local blueshifted temperature near the ECO surface is the Unruh
temperature
\begin{equation}
T_{s}=T_U(s)={1\over 2\pi s}\,.
\label{yfourt}
\end{equation}
The near surface region is filled with a thermal gas at temperature
$T(s)$, and if $T(s)=T_U(s)$ then the negative vacuum energy $\rho_v$
is cancelled by this thermal energy. In this case, we bypass the above
argument and the ECO can exist. Repeating this analysis with more
care, one finds that in 3+1 dimensions, if
\begin{equation}
s_{ECO}\ll \left ( {M\over m_p}\right)^{1\over 2}l_p\,,
\label{yten}
\end{equation}
then the ECO cannot exist if $T_{ECO}\ne T_H$. In $d+1$ dimensions,
this condition becomes $s_{ECO}\ll \left ( {M\over m_p}\right)^{2\over
(d+1)(d-2)}l_p$.

To make the above argument more precise, we write the metric outside
the ECO surface in the standard form
\begin{equation}
ds^2=-e^{2B(r)} dt^2+e^{-2A(r)} dr^2+r^2 d\Omega^2\,,
\end{equation}
and solve the Tolman-Oppenheimer-Volkoff equation in the region just
outside the ECO surface. We allow an arbitrary thermal stress-tensor
from thermal excitations which have $p_T(r)={1\over d}\rho_T(r)$ and from vacuum
energy, for which we take $p_v(r)={1\over d}\rho_v(r)$.\footnote{The tracelessness of the vacuum
stress tensor to leading order follows from the anomaly relation
giving the trace. The pressure $\rho_v$ is isotropic in the Boulware
vacuum, and we make the assumption that it continues to be isotropic
to leading order in the near surface geometry.} The approximation
$r\approx 2GM$ allows an explicit solution of the TOV equation, and
one finds that if (\ref{yten}) holds, then there is no solution where
$e^{A(r)},\, e^{B(r)}$ remain positive throughout the semiclassical
region outside the ECO.

This argument tells us that we need $T_{ECO}\approx T_H$, with the approximation becoming better and better as $s$ decreases below the scale on the RHS of (\ref{yten}). The general thermodynamic relation then tells us that
\be
dS_{ECO}\approx T^{-1} dE= (8\pi GM) dM\,,
\label{ytw}
\ee
 which integrates to
\be
S_{ECO}\approx 4\pi GM^2={A\over 4G}\,.
\label{ythir}
\ee
This steps (\ref{ytw}),(\ref{ythir}) were used to convert Bekenstein's qualitative conjecture $S\sim {A/G}$ \cite{bek} to the precise relation (\ref{ytwo}) after Hawking's discovery of the temperature (\ref{yone}); here we are just noting that {\it any} object with the same $T[M]$  as the black hole would yield the same entropy $S[M]$ as the black hole.

To obtain the radiation rate from an ECO, we first write the semiclassical computation of Hawking radiation in Schwarzschild coordinates. The effective potential $V_{eff}(r)$ for quantum field modes in the black hole geometry vanishes at $r\r 2GM$ and at $r\r \infty$, and has a `bump' in between. In the near horizon region, we have Rindler modes excited at the local Unruh temperature (\ref{yfourt}). These excitations have a small probability to tunnel through the `bump' in $V_{eff}$ and escape to infinity; the quanta that escape give Hawking radiation at the temperature (\ref{yone}). 

An ECO at temperature $T_{ECO}=T_H$ has radiation at the temperature (\ref{yfourt}) in the near surface region. Since the surface of the ECO is close to the horizon radius, the modes of quantum fields encounter virtually the same potential barrier $V_{eff}(r)$ as the modes in the black hole geometry. Thus, the spectrum of quanta escaping the barrier is the same, to leading order,  as the spectrum (\ref{ythree}) of radiation from the black hole.

In short, ECO's satisfying (\ref{yten}) must have the same thermodynamic properties (\ref{yone})-(\ref{ythree}) as the semiclassical black hole. 

How compact do we expect fuzzballs to be? In string theory, we have
seen that the count of microstates reproduces the Bekenstein entropy 
$A/4G$. Consider any theory of quantum gravity where the entropy is
order $A/G$.  Suppose we wish  to reproduce this entropy by having some
structure at the horizon. We will now give a heuristic argument which suggests that the object reproducing this entropy must be very compact \cite{ghm}.

Consider the quantum gravitational structure responsible for the entropy. We imagine this structure to be made of
Planck sized elementary objects arranged near the horizon. Each such object will cover an area
$\sim l_p^2$, so the number of such objects will be 
\be
N\sim \frac{A}{l_p^2}\sim  \left ({M\over m_p}\right )^2\,.
\ee
If we give each elementary object one bit of entropy (say, a spin which
can be up or down) then we recover an entropy $S \sim A/G$, as desired. But the mass of each elementary object will be $m\sim m_p$, giving a total mass 
\be
M_T\sim N m_p \sim  \left ({M\over m_p}\right ) M\,.
\ee
This is far larger than $M$.  How then can we think of the entropy in terms of structure at the horizon?

The key point is that if the microscopic structures mentioned above lie a short
distance $s$ outside the horizon radius $R=2GM$, then they sit at a location
with high redshift. Thus, the mass seen from infinity is
\be
M_\infty\sim (-g_{tt})^\h M_T\sim \left ({s\over GM}\right ) M_T \sim {s\over l_p} M\,.
\ee
We see that if $s\sim l_p$, then we get $M_\infty \sim M$, as required. Interestingly,  this  argument gives the same scale $s\sim l_p$ as the required location for the ECO surface in any spacetime dimension $d+1$.
 
Thus, on quite general grounds we expect fuzzballs to be very compact, with $s_{ECO}\sim l_p$. This length scale is much smaller than the limit (\ref{yten}), and so we expect the fuzzballs will reproduce the black hole thermodynamic relations (\ref{yone})-(\ref{ythree}) to an excellent approximation.

\section{Why is the semiclassical approximation violated?}

We have seen that in all  cases where we have been able to construct black hole microstates in string theory, these microstates have turned  out to be fuzzballs -- objects with no horizon. This resolves the information paradox, but at this point one might wonder: why did the classical picture of the hole change so radically in the full quantum gravity theory? The curvature at the horizon of a black hole of radius $R$ is ${\mathcal R}\sim {1\over R^2}$.  For a large black hole with $M\gg m_p$, this curvature  is very low: ${\mathcal R}\ll l_p^{-2}$. It is generally agreed that semiclassical physics should break down when ${\mathcal R}\gtrsim l_p^{-2}$. But here we are finding a large change all over the interior of the horizon, which is a region of low curvature at all points away from the central singularity. Thus, we are observing a second mode of failure of the semiclassical approximation. What is the trigger for this second mode of failure?

To answer this question, we will proceed in three steps. First, we will use rough estimates to understand why black hole dynamics can differ from dynamics in  other situations with low curvature. Next, we consider the quantum gravitational description of a static star, as we reduce its radius towards the horizon radius. We will see that in the limit, $R_{star}\r 2GM$ the virtual fluctuations of fuzzball type configurations become larger and larger. Finally, we will study the dynamics of gravitational collapse to understand why evolution to the semiclassical black hole geometry is not an allowed path in the quantum gravity theory, and the evolution leads to fuzzballs instead.

\subsection{The role of Bekenstein's  entropy}\label{sectunnel}

A key aspect in understanding the mysteries of the black hole  is the large value of the Bekenstein entropy (\ref{ytwo}). This entropy is far larger than the entropy of a gas with the same energy and volume as the hole. Now that we have some understanding of what this entropy represents -- it counts horizon sized quantum objects (fuzzballs)  that give the microstates of the hole -- we can ask what role these microstates play in the dynamics of gravitational collapse.

Consider a shell with energy $M$ that is collapsing to make a black hole. In semiclassical evolution, this shell passes smoothly through its horizon radius $R=2GM$, and continues inwards to create a singularity at $r=0$. But looking at the structure of the fuzzball geometries we know, we find that there is a path for the classical shell to tunnel into a fuzzball microstate. The tunneling amplitude will be very small, since we are talking about tunneling between two very different macroscopic  states. Let the spacetime dimension be $d+1$.   The probability of tunneling is
$P_{tunnel}=|{\mathcal A}|^2$
where ${\mathcal A}$ is the tunneling amplitude. In typical tunneling processes, we can estimate ${\mathcal A}$ as
\be
{\mathcal A}\sim  e^{-S_{cl}}\,,
\ee
where $S_{cl}$ is the classical action for the Euclidean gravitational path between the two configurations. To get some estimate of $S_{cl}$, let us set all length and time scales to be of order of the horizon radius $R\sim (GM)^{1\over d-2}$. This gives
\be
S_{cl} \sim {1\over G}\int d^{d+1} x \sqrt{-g} {\mathcal R} \sim {1\over G}R^{d+1} {1\over R^2} \sim \left ({R\over l_p}\right)^{d-1}\,.
\label{yttwo}
\ee
We see that the tunneling probability is indeed small for $R\gg l_p$. But the number of possible fuzzball states we can tunnel to is
\be
{\mathcal N}\sim e^{S_{bek}}\,,
\ee
where
\be
S_{bek}\sim {A\over G} \sim \left ({R\over l_p}\right)^{d-1}\,.
\ee
We see that it is possible to have
\be
P_{tunnel} \,  {\mathcal N}\sim 1\,.
\label{ytone}
\ee
so that the collapsing shell tunnels into fuzzballs in a short time, invalidating the semiclassical approximation \cite{tunnel}. In \cite{kraus} it was  argued that  the numerical coefficients in the exponents  in $P_{tunnel}$ and ${\mathcal N}$ are such that they actually cancel. In \cite{puhm} tunneling was considered between the members of a particular family of fuzzball microstates, and the enhancement due to the degeneracy factor analogous to ${\mathcal N}$ was observed.

 The argument leading to  (\ref{ytone}) is of course very heuristic; in particular, the simple estimate (\ref{yttwo}) may not give the tunneling rate to very complicated microstates. But (\ref{ytone}) does indicate that something special can happen for black holes that does not happen for a normal star: `entropy enhanced tunneling' can invalidate the semiclassical approximation even when the curvature scale is low. The larger we make the black hole, the smaller the curvature around the horizon, but at the same time the larger the Bekenstein entropy. Thus, the violation of the semiclassical approximation happens for all holes, however large  they are.  
 
 Put another way, in the path integral
\be
Z\sim \int D[g] e^{{i\over \hbar}S_{cl}}\,,
\ee
we usually assume that the measure factor is small for macroscopic systems, and thus we extremise the classical action $S_{cl}$ to get the semiclassical approximation. But the measure factor $D[g]$ involves the degeneracy of states that can contribute to the path integral, so for black holes we may write
\be
D[g]\sim e^{S_{bek}}\,.
\ee
The above estimates then show that the large value of $S_{bek}$ can make the measure term compete with the classical action,  thus invalidating the description of the black hole  as a semiclassical object.

\subsection{The VECRO hypothesis}\label{secvecro}

\begin{figure}
    \centering
    \includegraphics[width=0.6\linewidth]{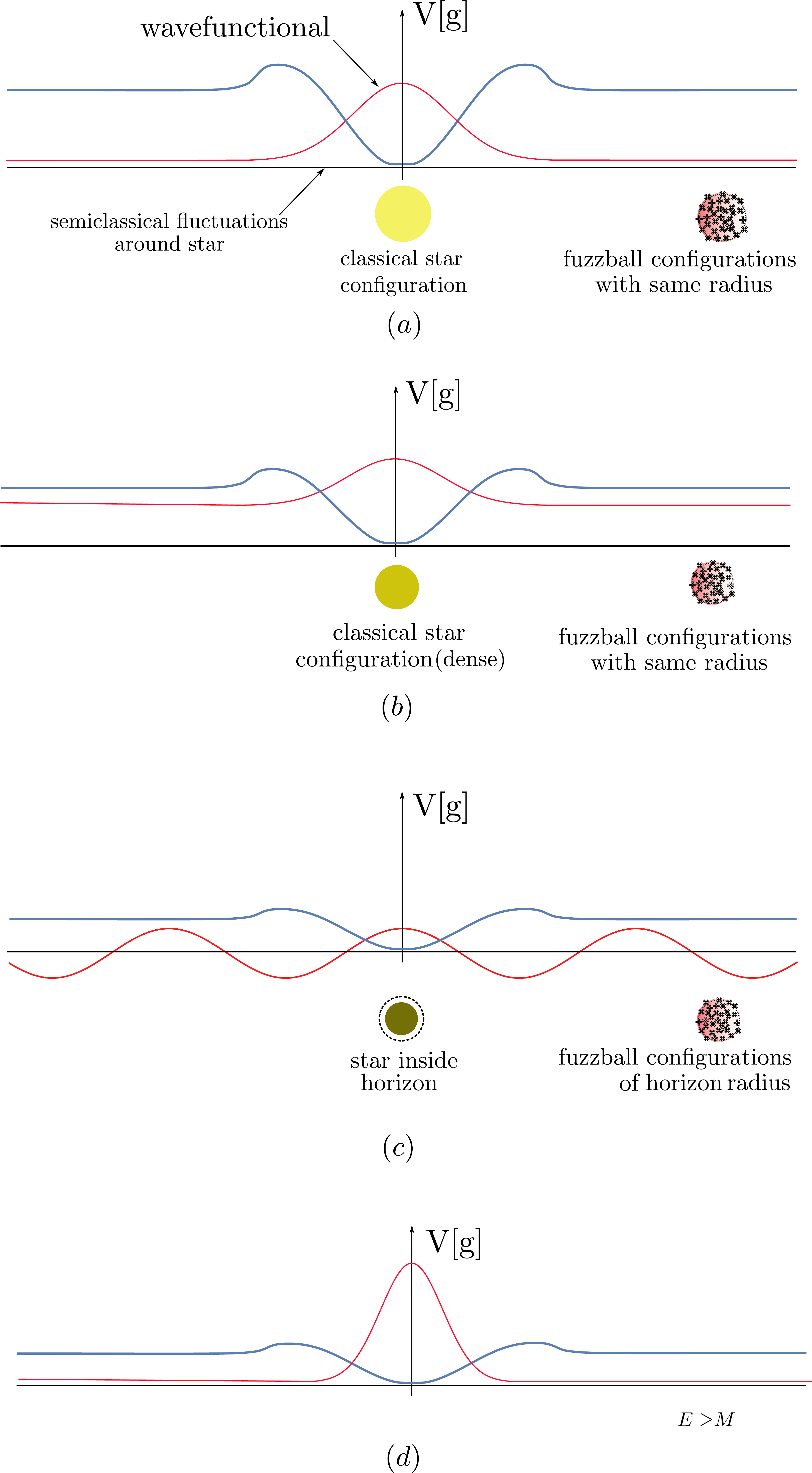}
    \caption{A schematic depiction of the potential function over the space of configurations of the gravitational theory. The thick blue line is the potential and the thin red line is the wavefunctional. (a) When a star is much bigger than its horizon radius, the fuzzballs with the same radius as the star are much heavier than the star, and the wavefunctional has a small support over these fuzzball configurations. (b) For a denser star, this support over fuzzball configurations is larger. (c) For a star inside its horizon radius, the wavefunctional spills over the entire space of fuzzball configurations of mass $M_{star}$. (d) For a star inside its horizon radius, we attempt to construct a wavefunction that is not allowed to spill over into the fuzzball region. The energy of the wavefunctional exceeds the mass of the star, so such a wavefunctional does  not describe a possible state of the collapsing star. }
    \label{fig:fuzzywavefunction}
\end{figure}

Let us now see how the above estimates can be used to develop a more explicit picture of gravitational wavefunctional. First consider a star of mass $M_{star}$ and radius $R_{star}$. In fig.\ref{fig:fuzzywavefunction}(a), along the horizontal axis, we depict schematically all the configurations of the gravity theory that are involved in describing the star. Thus, this is a space of large dimension, which for the sake of illustration, we are drawing as a 1-dimensional line.  The vertical axis gives the energy of the configuration. Thus, this graph is similar to the potential energy graph $V(x)$ for a quantum mechanics problem.

The central point on the horizontal axis is the classical configuration of the star. Around this point we have the small Gaussian deformations describing the quadratic fluctuations of quantum fields around the classical geometry of the star. Thus, this part of the potential graph is similar to the potential $V(x)=\h k x^2$ for a harmonic oscillator, except that there are many modes of the quantum fields with such a  quadratic potential. These quadratic fluctuations are the ones that lead to Hawking's particle creation under deformation of the geometry.
But further away from the configuration depicting the classical star, we must have the fuzzball configurations that describe the $Exp[S_{bek}]$ states that account for the black hole entropy. We focus on fuzzball states that have radius $R_{star}$, since the wavefunctional of the star can overlap with these states. Since the star has a mass $M_{star}<2GR_{star}$, these fuzzball states have a mass $M_{fuzz}$ much larger than the mass $M_{star}$ of the star. This fact is depicted in fig.\ref{fig:fuzzywavefunction}(a) by the fact that the potential graph in the region of the fuzzball states is high. In between the semiclassical configurations around the star and the fuzzball states, we have drawn a `bump'; this indicates that there is a barrier to tunneling between the star and the fuzzballs.

Now let us ask: what is the meaning of the evolution described by  classical gravity? The classical metric $g[x]$  depicts the peak of a quantum wavefunctional; the wavefunctional itself has a spread   around this classical peak. For our star, the wavefunctional is depicted by the dashed line in fig.\ref{fig:fuzzywavefunction}(a). Since the fuzzball configurations are also present in our space of gravity configurations, we must allow the wavefunctional to spread over these fuzzball configurations as well. Of course the potential is high at these fuzzball configurations, so the wavefunctional there is very small, and the semiclassical part describing Gaussian fluctuations around the classical configuration gives an adequate approximation to the wavefunctional.

Now consider the situation in fig.\ref{fig:fuzzywavefunction}(b). The star still has mass  $M_{star}$, but is denser. The mass of the fuzzballs with radius $R_{star}$ is still more than $M_{star}$ but not  too much more. The potential in the fuzzball region is now lower, and the wavefunctional has a larger amplitude over these fuzzball configurations.

In fig.\ref{fig:fuzzywavefunction}(c) we  attempt to shrink the star inside its horizon radius. Now, the wavefunction does not stay confined to the potential well around the classical configurations; it spreads over the vast space of fuzzball configurations. For a toy model of this phenomenon, consider the quantum mechanics problem of a square well of depth $V_0$ and radius $a$,  in a large number of dimensions $d_n$. For a particle of mass $m$, there is a bound state in this well only if
\be
V_0\gtrsim {\hbar^2\over 8ma^2} d_n^2
\ee
If $V_0\lesssim {\hbar^2\over 8ma^2} d_n^2$, then the  wavefunction `pops out' of the well and spreads over  the entire spatial region. This is what has happened to our star: while classical gravity would suggest that a star can shrink inside its horizon, we must consider if this classical configuration can really be the peak of a wavefunctional which is localized around this classical configuration. The large dimension of the space of fuzzballs (given through the Bekenstein entropy) makes it impossible to trap the wavefunctional into the Gaussian part of the potential well when the radius of the star reaches $R_{star}\approx 2GM_{star}$, since all the $Exp[S_{bek}]$ states of mass $M_{star}$ are accessible to the wavefunctional. Thus the semiclassical approximation breaks down when the star tries to shrink inside its horizon radius.

Finally, in fig.\ref{fig:fuzzywavefunction}(d) we depict a state where we force the wavefunctional to be localized in the Gaussian part of the potential even though the  star  has collapsed through its horizon. While we can certainly consider such a wavefunctional, it has an energy $\langle E\rangle$ which is higher than $M_{star}$; thus it does not describe a star of mass $M_{star}$ which has collapsed through its horizon.  The situation can again be understood with the help of our toy model of the potential well. For, $V_0\lesssim {\hbar^2\over 8ma^2} d_n^2$ we can  certainly consider a wavefunction that is localized inside the well.  But this wavefunction will not be a bound state with $E<0$; it will have an expectation value of energy that is higher than the continuum, i.e., we will find $\langle E\rangle >0$.

To summarize, classical evolution describes the peak of a wavefunctional that in principle spreads over all quantum states. When we try to shrink a star inside its horizon, the wavefunctional does not stay narrowly localized around this classical configuration, so the semiclassical approximation breaks down.

The notion that the fuzzball component of the wavefunctional is important is called the VECRO hypothesis \cite{vecro}. The term VECRO stands for Virtual Extended Compression-Resistant Objects. The fuzzball tail of the semiclassical wavefunctional consists of `virtual' fluctuations of `extended' objects, since the fuzzball configurations are not pointlike particles but rather large complex configurations.  As we will see below, these configurations are resistant to compression or extension, a property that is important in their contribution to gravitational dynamics. We will now see the role of the vecro part of the gravitational wavefunctional in the process of black hole formation. 

\subsection{The dynamics of gravitational collapse}\label{secdetail}

Finally, we come to the crucial question: why does gravitational collapse not lead to the semiclassical geometry of the hole? Consider a shell of mass $M$ composed of radially ingoing massless quanta, in 3+1 dimensional asymptotically flat Minkowski space. By causality, no particle on this shell can receive a signal from any other particle on the shell. Thus, it would seem that each particle will move independently through empty space,  crossing the horizon uneventfully and progressing towards the center. New physics can certainly arise at the center since ${\mathcal R}\gtrsim l_p^{-2}$ at the singularity, but by causality this new physics cannot affect the horizon which is spacelike separated from the singularity. It would therefore seem that we will have a semiclassical horizon and the associated information loss problem. What is the way out?

As we will now see, the particles of the shell indeed cannot exchange signals with each other. But once the shell reaches the vicinity of the horizon,  the particles of the shell do not leave behind them a vacuum spacetime.  In the region of this `wake' (i.e., the region $r>R_{shell}$)  we {\it can} receive information from a curved segment of the shell. If the shell is at a radius $R_{shell}\gg 2M$, then this `wake' region is close to the vacuum,  as expected from usual semiclassical dynamics. But for $R_{shell}\approx 2GM$, it is important to note that  the metric outside the shell is  a Schwarzschild metric with mass $M$.  We had noted in section \ref{secvecro} that the fluctuations of virtual fuzzballs -- the vecro component of the wavefunctional -- will be larger and larger as the shell approaches the horizon. When the shell crosses into the horizon region, the wake left behind the shell generates fuzzballs rather than the traditional vacuum geometry.

To arrive at this picture of evolution, we need to provide a crucial ingredient: a reason why the `wake' region behind the shell cannot just be the semiclassical vacuum.  We will now give a bit model of the gravitational vacuum which incorporates the vecro hypothesis; with this model  we will  find that the collapsing shell cannot give the semiclassical black hole geometry but must yield fuzzballs instead. We proceed in the following steps:

\b

(A) We know that the vacuum of quantum theory has fluctuations of electrons and positrons; it is these fluctuations that turn to on-shell particles in the process of Hawking radiation. But the electron and positron also form a bound state -- the positronium. Is there any effect of this on-shell bound state on the structure of the  vacuum? The answer is yes: the fluctuations where the electron and positron are at a separation of order the positronium radius have a slightly greater amplitude than fluctuations where they do not have this separation.  Thus, bound states in the theory lead to correlations among the vacuum fluctuations.  The particles of the standard model have only a few bound states at any given energy, and as we  consider bound states of  higher and higher energies, their fluctuations are quite suppressed anyway. Thus, it may seem that the correlations induced by bound states are not an important aspect of the vacuum wavefunctional. 

But could there be some large class of bound states that we have conventionally ignored? Yes: the $Exp[S_{bek}(M)]$ microstates of black holes for each mass $M$! It is true that the virtual fluctuations corresponding to  states of larger mass $M$ will be suppressed, but we have already noted in section \ref{sectunnel} that this suppression can be offset by the larger degeneracy of bound states at larger $M$. The vecro hypothesis says that there are important correlations in the vacuum at all length scales due to the presence of on-shell black hole microstates in the theory. We will now make a model of these correlations, and see what they do in the process of gravitational collapse.

\b

\begin{figure}
    \centering
    \includegraphics[width=\linewidth]{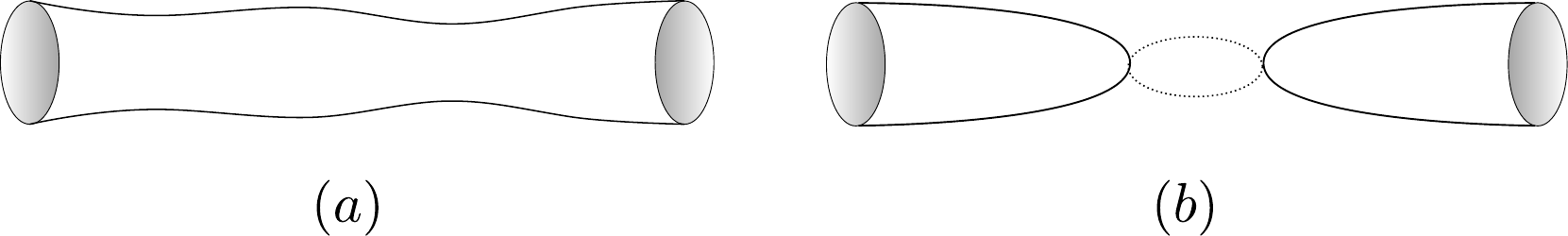}
    \caption{Small fluctuations of a compact circle give a massless scalar in the low energy dimensionally reduced theory. (b) At high energy scales we have larger fluctuations that create topological deformations like KK monopoles and anti-monopoles; these are the bits carrying spin that we use in fig.\ref{fig:spins} below. }
    \label{figkk}
\end{figure}

(B) It is generally agreed that spacetime has violent quantum fluctuations at the Planck scale. Let us make a toy model of these fluctuations; this will enable us to talk about correlations among these fluctuations in a more concrete way.  Consider a spacetime with a compact circle, depicted in fig.\ref{figkk}(a). Small fluctuations in the radius of this circle describe a massless scalar in the dimensionally reduced theory. But we can also have larger fluctuations which pinch off the circle and create, for example, a KK monopole or antimonopole (fig.\ref{figkk}(b)).  In string theory, the KK monopole forms a 256 dimensional multiplet after we take into account fermion zero modes localized at the monopole. Here we will just assume for simplicity that at the microscopic scale we have excitations (the analogue of KK monopoles) that carry a spin $\h$ degree of freedom. We can thus imagine a lattice of spin $\h$ bits, with the proviso that as space expands  in volume we can create more  lattice sites and when space shrinks in volume we move to a configuration with less lattice sites.
\b

(C)  The vacuum is the state with the lowest energy.  Because of the Hamiltonian interaction between neighboring spins, the lowest energy state of the theory will be such that the spin at any site is entangled with the spins at nearby sites. This entanglement will capture the correlations that we seek to model.  The correlations result from the presence of  virtual fuzzballs with all possible sizes. Thus, we should have correlations at all length scales, though the correlations should fall  off with distance. Let us make a toy model describing these correlations.
\begin{figure}
    \centering
    \includegraphics[width=\linewidth]{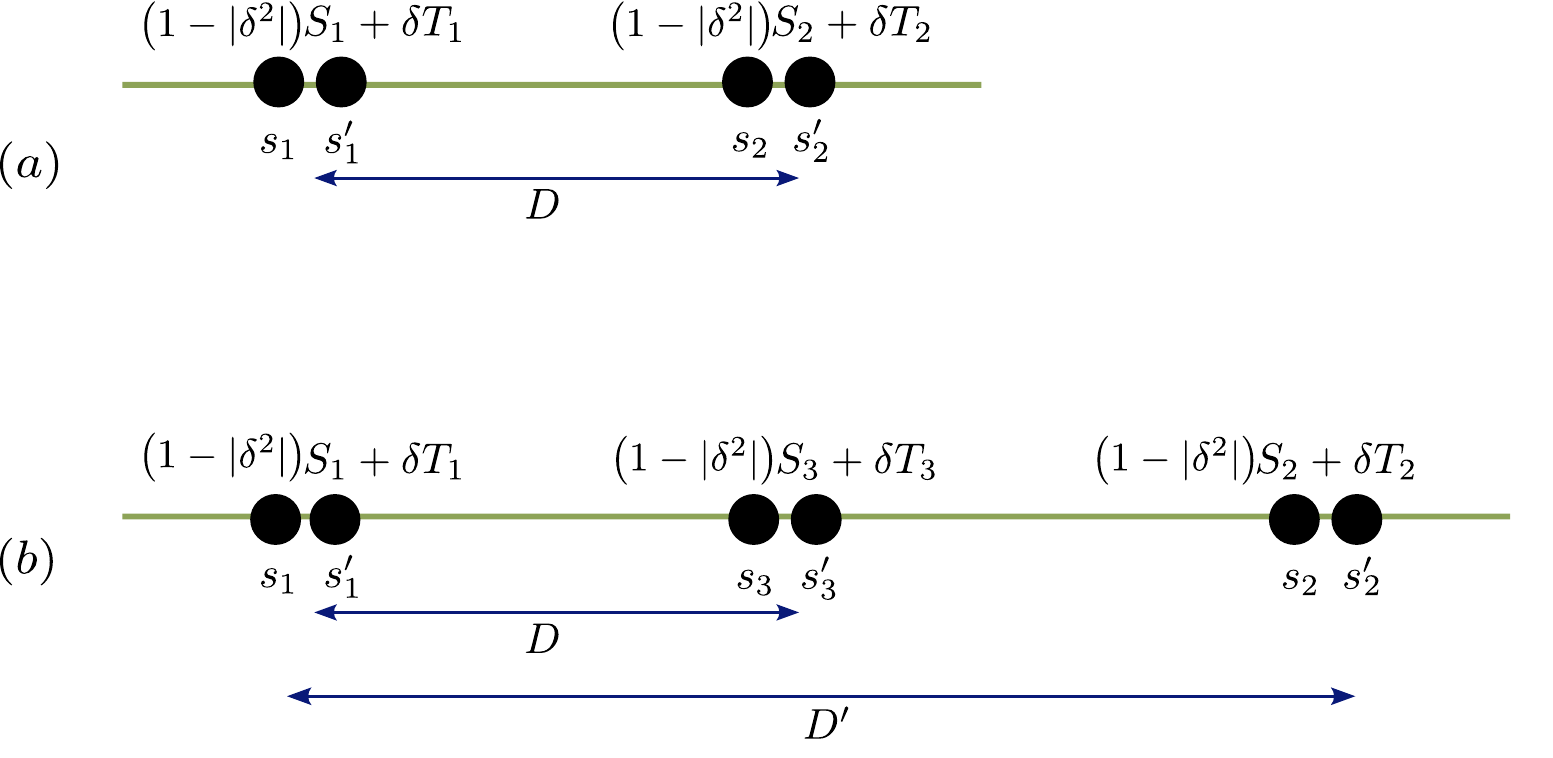}
    \caption{(a) Neighboring spins $s_1s'_1$ form a state that is mostly a singlet $S_1$ but with a small admixture of the triplet $T_1$, and similarly for $s_2s'_2$. Entangling the triplets into a singlet gives a situation where each spin is largely entangled with its nearest neighbor, but has a small entanglement with spins further away. (b) When space expands, we create a new spin pair $s_3s'_3$. For the stretched space to be a vacuum, we must have the same entanglement structure at separation $D$ as we had in the initial manifold in (a). }
    \label{fig:spins}
\end{figure}
First, suppose that we had correlations only among neighboring sites. (These will be correlations arising from the virtual fluctuations  of the smallest black holes.)  We  group the spins in pairs. Consider a pair of neighboring spins called $s_1s'_1$. Let their singlet state be $S_1$ and triplet $T_1$. If we only wished to consider the entanglement arising from the smallest virtual black holes, then we would let the state of this pair be $S_1$.

Now let us see how we can also allow correlations across longer distances, which would arise from virtual fluctuations of large black hole microstates. We want an entanglement that is large between neighboring spins, and smaller  (but not zero) between spins that are further away. We can model this as follows. For the two neighboring spins $s_1s'_1$ considered above, we take the combined state to be mostly the singlet $S_1$, but with a small admixture of the triplet $T_1$
\be
|\psi\rangle_{12}=(1-|\delta|^2)S_1+\delta  \, T_1\,, ~~~\delta  \ll 1\,.
\ee

Consider another similar pair of spins $s_2s'_2$ that are at a distance $D$ away, and which form a similar spin state: mostly the singlet $S_2$ and a bit of the triplet $T_2$. We then entangle the triplet states $T_1, T_2$ between these two distant pairs of spins to make a singlet. This is a toy model describing a situation where we have a large correlation with nearby spins, as well as a small but nonzero correlations between spins that are further away. By making such a hierarchical set of entanglements between a given spin and other spins at various distances, we can include the effect of vecros (virtual black hole microstates) of all sizes.

\b

(D) Let us now input the energy scales of black hole physics into our model. We have a hierarchical set of correlations. There is an entanglement between the spins in the smallest block of radius $2$ lattice sites. Next we group these blocks into larger blocks of radius $4$ lattice sites; these blocks in turn are grouped into blocks with radius $8$ lattice sites and so on. At each stage of the blocking, we have an entanglement between the blocks that are joining to make the next larger block, of the kind described in (C) above. Consider the blocks of radius $R_v$, which will correlate with each other to make the next larger block of radius $2 R_v$ (the symbol $R_v$ stands vecro radius). These correlations  of the regions of radius $R_v$ contribute a (negative) energy, $E(R_v)$ which we set to be of order of the black hole mass for radius $R_v$. In 3+1 dimensions, this will be
\be
E(R_v)\sim -{R_v\over G}\,.
\ee
In terms of energy density, this is 
\be
\rho(R_v)\sim -{1\over R_v^3} E(R_v) \sim -{1\over G R_v^2}\,.
\label{bbqone}
\ee
To summarize, the virtual black hole microstates (vecros) of radius $\sim R_v$ lead to correlations in the vacuum that lower the energy density by the amount (\ref{bbqone}). Taking into account the lowering of energy density by all vecros with $0<R_v<\infty$ brings us down to the net vacuum energy density which is assumed to be zero since we are starting with flat Minkowski space. From all this, we see that if  we have a state with the correct correlations  at scales $0<R_v<R_v^{max}$ but not at scales $R_v>R_v^{max}$, then we will get a {\it positive} energy density of order 
\be
\rho\sim {1\over G (R^{max}_v)^2}\,.
\label{bbqtwo}
\ee

\b

(E) Now we come to a crucial point. Suppose the space depicted in fig.\ref{fig:spins} expands. The two pairs of spins $s_1s'_1$ and $s_2 s'_2$ get separated to a distance $D'>D$,  and new spins appear at lattice points in between. To get the optimal correlations that lead to the zero energy vacuum, we will need the triplet $T_1$  of the first pair to entangle with the triplet $T_3$ of the new pair $s_3s'_3$ that is a separation $D$ away in the new metric. But now we see an important issue. For the triplet $T_1$  of the first pair to entangle with the triplet $T_3$ of the new pair, we must first remove the entanglement between the two original triplets $T_1, T_2$. We can remove this entanglement by suitable interactions between the two sets of spins $s_1s'_1$ and $s_2s'_2$, but for these interactions to take place we must have enough time for light to travel between  $s_1s'_1$ and $s_2s'_2$.  We now have two cases:

\b

(i) If the expansion is slow, light will be able to traverse between  $s_1s'_1$ and $s_2s'_2$ several times in the course of the expansion. Thus, we can  reach the entanglement appropriate to the vacuum in the new expanded lattice; i.e.,  $T_1$ will disentangle from $T_2$ and entangle with $T_3$. Under this adiabatic expansion, the initial zero energy vacuum stretches to the new zero energy vacuum.

\b

(ii) If the expansion is rapid, then light cannot travel between  $s_1s'_1$ and $s_2s'_2$ during the expansion. Then we cannot remove the entanglement between, $T_1,T_2$ and thus we cannot entangle $T_1,T_3$. We do not reach the vacuum state of the expanded manifold. Instead, we get a state with extra energy density of order (\ref{bbqtwo}), where $R_v^{max}$ is given by the distance light has been able to travel during the expansion. 

\b

Note that the two cases of adiabatic and nonadiabatic expansion also arise in the usual study of quantum fields on curved space. What is new here is that there are extra correlations in the vacuum that stretch across all length scales; these correlations arise from virtual black hole microstates (vecros) which are {\it extended} objects spanning all length scales. The contribution of these correlations is significant because of the large degeneracy of bound states implied by the Bekenstein entropy.  

\b

(F) With all this, we can now explain how the semiclassical approximation can break down in black hole formation but not in other places where we have seen general relativity to be valid:

\b

\begin{figure}
    \centering
    \includegraphics[width=0.6\linewidth]{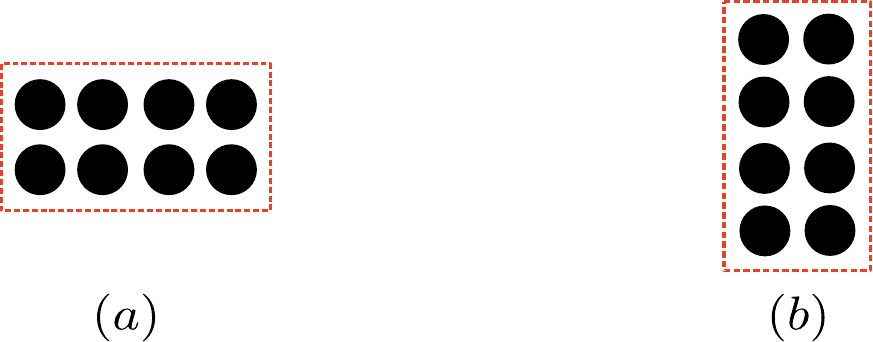}
    \caption{In a gravitational wave, the spin carrying bits of fig.\ref{fig:spins} just rearrange themselves locally to create a shape distortion from (a) to (b); we do not need to create or annihilate bits. Thus, the dynamics of vecros is not involved in this situation, even though the wave moves fast (at the speed of light).  }
    \label{fig:gravwave}
\end{figure}

(i) In a gravitational wave, there is no increase in the volume of space; we just deform its shape. Consider a wave traveling in the $z$ direction. When the $x$ direction is compressed, the $y$ direction  expands and vice versa. In fig.\ref{fig:gravwave} we see how the microscopic bits can rearrange themselves locally to create this distortion. This distortion does not change overall volume and thus not require the creation or annihilation of bits. Since this rearrangement of bits is at the local level, it can be completed in order of Planck time, which is much smaller than the period of the wave. We do not have any issue of the kind described in (E)(ii) above. Thus, we do not get any novel effect from vecro dynamics, and general relativity is a valid approximation for waves with $\lambda \gg l_p$.
\b

(ii) Consider the formation of a star from a ball of gas.  The  particles move much slower than the speed of light, and so light has time to traverse the star multiple times during the star-formation process. The entanglements required to minimize the vacuum energy can be created, and general relativity is again a valid approximation.

\b

\begin{figure}
    \centering
    \includegraphics[width=0.6\linewidth]{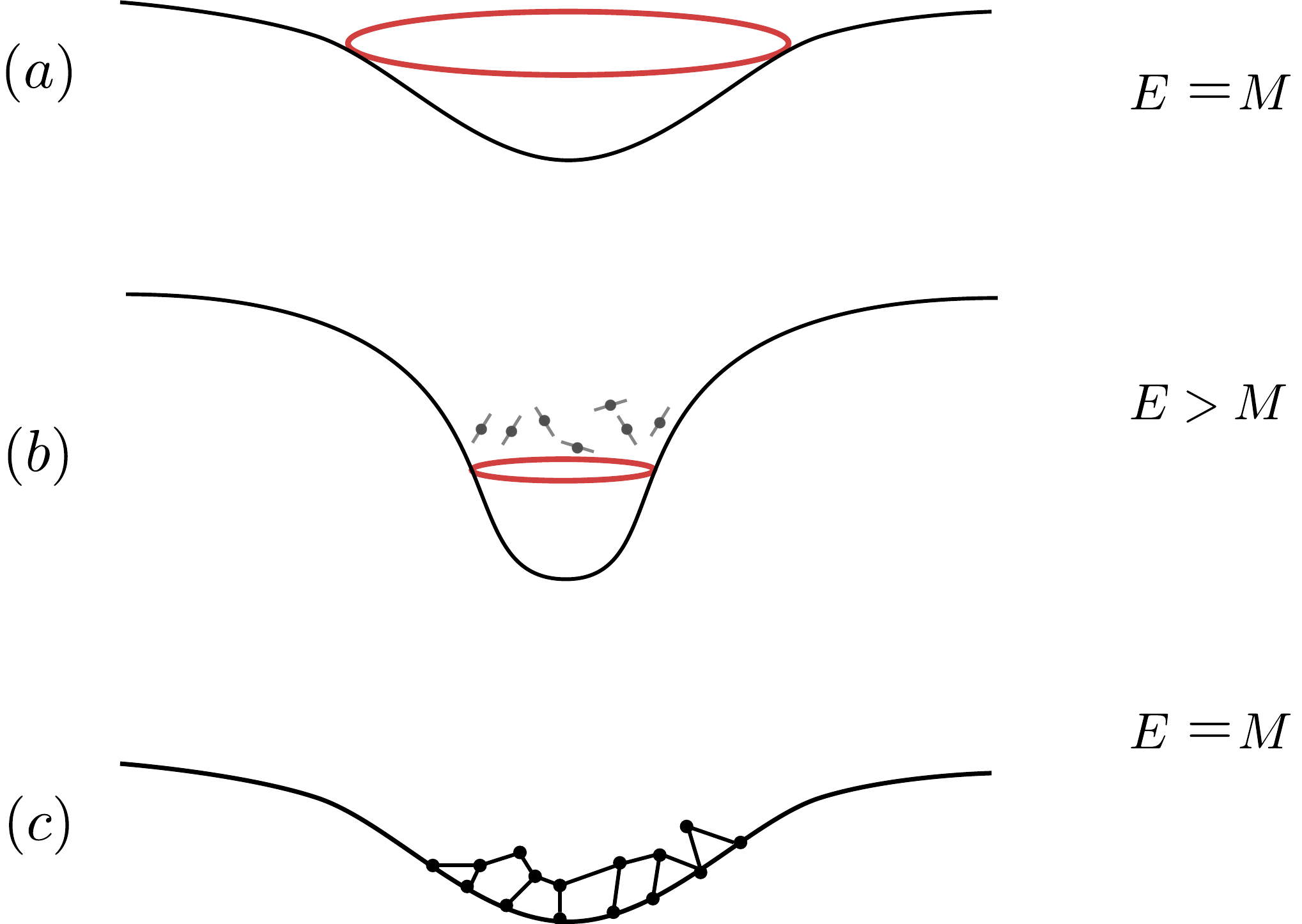}
    \caption{(a) The good slicing of the semiclassical hole manifests a stretching of spatial slices as a shell (indicated by the thick red circle) comes in from infinity. (b) Light signals cannot travel between different points of this stretching region to establish the vecro correlations appropriate for the quantum gravitational vacuum wavefunctional. As a result we have local excitations (indicated by dark dots) that are entangled with each other, in the manner of fig.\ref{fig:spins} (b), and the configuration is `under the potential barrier' with energy $E>M$ (c). The spins in the dark dots can link up in a fuzzball configuration  which has energy $E=M$; this is the result of tunneling from the collapsing shell state to a linear combination of fuzzballs. }
    \label{fig:shell}
\end{figure}

(iii) Consider a shell that is collapsing at the speed of light to make a black hole.  The black hole puzzle appears strongest when seen in a `good slicing', so we draw such slices in fig.\ref{fig:shell}. We see that the slices stretch progressively as the time at infinity evolves. But note that inside the horizon, the analogues of the spins $s_1s'_1$ in fig.\ref{fig:spins} cannot communicate with the analogues of spins $s_2s'_2$ on the diametrically opposite side, due to the inward pointing structure of light cones. Spins inside the horizon also cannot send signals to the spins just outside. Thus, we will not be able to establish the correlations appropriate to the vacuum spacetime over distances of size $\sim GM$. Thus, in (\ref{bbqtwo}) we set $R_v^{max}\sim GM$. Using this $\rho$ and a volume $\sim (GM)^3$ for the black hole region, we find an extra energy from the vecro effect on the vacuum
\be
E_{extra}\sim \rho (GM)^3 \sim {1\over G (GM)^2} (GM)^3\sim M\,.
\ee
This energy $E_{extra}$ was missing in a semiclassical analysis of collapse. We see that the classical black hole spacetime with stretching slices cannot be a valid on-shell solution of the full quantum gravity theory. The configurations of fig.\ref{fig:shell}(b) cost more energy than we have available. This extra energy is carried by virtual excitations of bubbles --- KK monopoles and antimonopoles in our toy model --- which are separated spatially but have their spins entangled with each other. In a quantum mechanical analogue, these configurations would  correspond to points where a particle is `under a potential barrier'. In fig.\ref{fig:shell}(c) we show configurations where the  bubbles have linked up into a configuration that describes a fuzzball of mass $M$; transitioning to one of these configurations is like emerging from under the potential barrier to a new set of allowed states. This is the process of tunneling to fuzzballs. 

\b

(G) We have seen above that  semiclassical physics breaks down when `space stretches too fast'. Where else do we find such fast stretching? Consider cosmological expansion. The spatial volume doubles in a time $\sim H^{-1}$, where $H$ is the Hubble constant.  If we consider a region with size $L\ll H^{-1}$, then light can traverse this region several times during the doubling time. As we noted above, in this situation the vacuum correlations at scales $L$ will settle down to their vacuum values, and we will get no novel effects from the vecro part of the quantum gravity wavefunctional. But consider regions of size $L\sim H^{-1}$. Here light has barely enough time to cross the region in doubling time, and according to our vecro hypothesis, we expect extra energy density (\ref{bbqtwo}) with $R_v^{max}\sim H^{-1}$. In \cite{elastic} it was noted that this extra energy might help resolve issues like the Hubble tension, the cosmological constant, and the energy needed to drive inflation; all of these are effects involving a mysterious source of energy density of the order given by (\ref{bbqtwo}). 

\subsection{Summary}

In this section, we have seen how  semiclassical evolution breaks down in the process of black hole formation, even though the curvatures are low in the region around the horizon. This is a second mode of failure of the semiclassical approximation, different from the usual mode where ${\mathcal R}\gtrsim l_p^{-2}$. This second mode involves fast stretching of space, which leads to a non-adiabatic evolution of correlations that arise from virtual black hole microstates. We always knew that the Bekenstein entropy was large, but without an understanding of the microstates counted by this entropy, we could not find the role of this entropy in black hole dynamics. With the fuzzball construction, we now understand that black hole microstates are extended structures of horizon size. The corresponding correlations in the vacuum wavefunctional -- vecros -- therefore also extend over all length scales, and create extra energy when the expansion is too quick to allow these correlations to settle down to their optimal form \cite{secret}. This extra energy creates a potential barrier to evolving along the traditional semiclassical path, and leads to a tunneling to fuzzballs instead.  This is the resolution of the information paradox. 

Note that we have maintained causality and locality at all steps in our analysis: no signals propagate outside the light cone, and the Hamiltonian of the fundamental theory does not have interactions between spatially distant points. The vecro correlations on all length scales in the Minkowski vacuum arise from this local Hamiltonian, and can exist because this space has an infinitely long past history which allowed these correlations to form. In an expanding cosmology, light has not had time to travel further than the particle horizon, and so there are no vecro correlations outside this distance; this was an essential ingredient in the cosmological analysis of \cite{elastic}. 

Correlations across spatial distances also arise in the local, causal theory of a scalar field in Minkowski space, and are reflected in the Wightman functions between spatially separated points. One might then ask if  vecro correlations can be seen in quantum fluctuations of the observed low energy quantum fields of the standard model. But in string theory, the  standard model fields arise as the analogues of the gravitational waves depicted in fig.\ref{fig:gravwave}; the only (irrelevant) difference is that the polarization of these waves is in the compact directions.  As we saw in section\,\ref{secdetail}, part (F)(i), in a gravitational wave we do not  change  the number of bits involved in the gravitational wavefunctional; we simply rearrange these bits locally. Thus while vecro correlations are in principle just like the correlations across spacelike distances observed in usual quantum fields, they involve a different set of excitations, and affect dynamics only when we have the `fast stretching' described above.\footnote{In a rough analogy, we may say that the usual low energy quanta of the standard model are like phonons in a superfluid, while the bits involved in the vecro correlations are like the rotons.}

\subsection{The possibility of fuzzball complementarity}

We still have one question left: what is the fate of an infalling observer?  In the classical black hole, the observer would pass uneventfully through the horizon, and feel strong tidal forces only as he approaches the singularity at $r=0$. In the fuzzball paradigm, the observer would transition to fuzzball degrees of freedom as he reaches the horizon. Is there still some way that he could feel that he is falling through the horizon into a smooth black hole interior?

To understand the answer, it is useful to step back into the history of different ideas and counter-ideas that have been proposed for  this `infall'  question.

Starting around 1988, 't Hooft  argued in a set of papers that matter falling in towards the horizon of a black hole  will bounce off the horizon and return to infinity \cite{thooft}. It was soon noted that his scattering process used `post-selection' of states, and was differed from the  usual Schr\"odinger evolution along the `good slices' of the black hole geometry.

't Hooft's argument suggested that ingoing and outgoing modes were not different parts of a quantum field, but were `complementary' in the way that position and momentum are complementary in quantum mechanics. Susskind et al.  \cite{susskindetal}  attempted to develop the idea of complementarity with the following postulates. (i) Infalling matter gets absorbed by a surface (called the stretched horizon) just outside $r=2GM$, and the information in this matter gets radiated  back to infinity from this surface.(ii) There is a  `complementary' description in which the infalling matter encounters nothing at the horizon and falls smoothly through.(iii) The two copies thus created --- one inside and one outside --- appear to violate the `no cloning' theorem of quantum mechanics. But it is argued that `no cloning' can be violated here, since an observer cannot measure one copy and then the other because of the constraints presented by the horizon.

It was not clear what to make of this notion of black hole complementarity; it seemed to require postulating new physical principles for which we had no evidence. In particular, it was not clear what mechanism would scramble and return information at the stretched horizon, since at the time there was no evidence of any degrees of freedom that could stand at this location without falling through the horizon. But this proposal of complementarity suggested in an abstract way the notion of holography, where the degrees of freedom of the hole could be regarded as living at the boundary of the hole; i.e., at the horizon.

The years 1995 and 1996 saw rapid progress in understanding black hole entropy and the structure of black holes in terms of bound states of D-branes. In June 1997 it was estimated that the size of these bound states was always of the order of the horizon radius, suggesting that black holes in string theory are normal objects with no horizon \cite{emission}. Further studies of brane bound states continued to support this conclusion, leading to the emergence of the fuzzball paradigm discussed  in this article \cite{fuzzballreview, bwreview, borunreview}. The fuzzball resolution of the information paradox is quite different from  Susskind's conjecture of complementarity. Susskind had required that the horizon appear to be a local vacuum in some choice of variables.  But  with a fuzzball, the low energy modes involved in Hawking radiation do not see a vacuum horizon in any choice of variables.   

In November 1997, a concrete realization of holography was obtained in string theory by Maldacena \cite{adscft,gkp,wittenads}. Imagine a graviton falling onto a stack of D3 branes. The graviton gets absorbed by the branes, its energy getting converted to a complicated collection of open strings stretching between the branes. But with a change of variables, these open strings can be written in terms of closed strings. In this dual  description, the infalling graviton passes smoothly into an AdS space, encountering no branes in the process. In the open string description -- the `CFT' -- the ball of open string excitations spreads with time  over the surface of the D-branes. In the dual `gravity' description, we simply see the graviton fall deeper into empty AdS space. This  AdS/CFT duality is a realization of holography, since in the CFT description the degrees of freedom live only on the surface of the branes, while in the gravity description we have an extra radial direction orthogonal to these branes  in which the graviton moves. 

The conjecture of AdS/CFT duality says that all the gravitational physics inside AdS space can be recast, in different variables, as a field theory without gravity. We can make and evaporate a black hole in AdS space. Since the field theory is unitary, this indicates that black hole evaporation should be unitary as well. But the AdS/CFT conjecture  does not tell us the resolution of the information paradox, since it leaves all  possibilities open: (i) The black hole microstates are fuzzballs, so there is no pair creation at a  horizon (ii) The hole has its semiclassical form, but nonlocal effects (e.g., wormholes) somehow transport information from the singularity to the radiation region (iii) AdS/CFT duality itself breaks down or gets modified when a black hole forms; after all, quantum theory is valid everywhere other than in the black hole Hawking process, so a similar situation could hold for AdS/CFT duality.  

Given this, how should we relate the AdS/CFT duality conjecture to black holes? When this duality was initially proposed, the fuzzball idea was still quite new, and many people considered it too radical a change to the structure of the hole. They assumed that in the gravity description, black holes would continue to be described by the semiclassical hole, at least to a first approximation. But then, how do we evade the Hawking puzzle of monotonically growing entanglement?

One suggestion was that the Hawking computation was itself flawed. Hawking did a leading order computation of pair production using quantum fields on curved space. But there can always be small (hitherto unknown) quantum gravity corrections (say $O(\epsilon), \epsilon\ll 1$) to this leading order computation. The number of quanta $N$ emitted by the hole is large, with $N\sim (M/m_p)^2$. Even though the correction to the state of each emitted pair would be small, these small quantum gravity effects could induce subtle correlations between the large number of quanta $N$. If
\be
\epsilon \, N\gg 1\,,
\ee
then the large number of these `small correction' terms might overwhelm the leading order computation and yield an entanglement $S_{ent}=0$ for the radiation at the endpoint of evaporation. Then there would be no information paradox.

In 2004, Hawing conceded that his argument for information loss might be invalid due to such small corrections \cite{hawkingretract}. He noted that the Euclidean path integral would have subleading saddle points. A subleading saddle would produce small corrections to Lorentzian evolution, and perhaps the subtle correlations induced thereby would remove the troublesome entanglement.

But most relativists were perplexed by Hawking's retraction. If the correction to each pair is small, then they expected that the correction to the overall entanglement would be small; thus, such small corrections should have no bearing on the original argument of information loss.

In 2009, it was shown that such small corrections {\it cannot} remove the troublesome entanglement \cite{cern}. If the correction to each pair is small as in (\ref{utwo}), and if there are no nonlocal Hamiltonian interactions linking the black hole to the radiation at infinity, then the entanglement must continue to rise monotonically like (\ref{uone}). This `small corrections theorem' established fuzzballs as the natural resolution of the information paradox: the theorem says that order unity corrections are required at the horizon to resolve the Hawking puzzle,  and all the microstates constructed so far in string theory are found to be fuzzballs which have no horizon. 

We have noted above that the conjecture of AdS/CFT duality cannot by itself tell us how to resolve the information paradox. But this duality conjecture did suggest a possibility for the {\it infall} problem. Hawking radiation consists of quanta with energies $E\sim T$, where $T$ is the temperature of the hole. These radiated quanta cannot see a vacuum around the horizon in any set of variables, since quanta emerging from the vacuum have the state (\ref{utwo}) and not a state that carries information about the hole.  But the infall problem addresses observers that fall {\it into} the hole. The temperature of the hole is very low:  the wavelength $\lambda\sim 1/T$ is order the horizon radius. Thus, we can focus on infalling objects which have $E\gg T$. Could it be that in some choice of variables, such infalling objects feel that they are falling through the classical black hole geometry, to a first approximation?  This experience of free fall would have to be approximate, since it must fail completely when $E\sim T$; thus the best that we can hope for is that the approximation becomes better and better in the limit $E/T \gg 1$. 

More explicitly, the analogy with AdS/CFT would be as follows:

(i) The infalling object reaches the surface of the hole, and its energy gets converted to fuzzball degrees of freedom, yielding a complicated stringy mess. The analogue in AdS/CFT would be that a graviton hits a collection of D3 branes and gets converted to a messy collection of open strings. 

(ii) In some other variables, the object falling onto the horizon  would appear to continue through the horizon into a black hole interior. The analogue in AdS/CFT would be the gravity description, where the incoming graviton continues to fall smoothly into AdS space.

The black hole case would still differ from AdS/CFT in the following way: the dual smooth infall (ii) emerges only as an approximation that works when $E\gg T$, and fails when, we consider the modes with $E\sim T$ that are involved in Hawking radiation.

The above  possibility for infall into a black hole is called `fuzzball complementarity'. The word `complementarity' is meant to acknowledge similarities with Susskind's notion of complementarity mentioned above. The word `fuzzball' emphasizes the difference from Susskind's complementarity. Susskind had imagined that a vacuum region around the horizon would emerge in some variables, and all modes including those with $E\sim T$ could be studied in these variables. In fuzzball complementarity, the modes with $E\sim T$ do not see a vacuum region in {\it any} choice of variables, and complementarity emerges only as an approximation for $E\gg T$.

In 2012, Almheiri, Marolf, Polchinski and Sully (AMPS) gave a `firewall' argument to show that Susskind's proposal of complementarity could not work \cite{amps}. In the variables described by the stretched horizon, the hole is supposed to radiate like a normal body. For a normal body, in the second half of the evaporation process,  the degrees of freedom in the body are almost maximally entangled with the emitted radiation, since the body has fewer bits than the radiation.  Consider a hole that is in this second half of its evaporation history. Suppose that in some variables we obtain a horizon  with a vacuum region around it, as required by Susskind's complementarity. Then the bits in this vacuum region must be entangled with each other in the form (\ref{bits}); i.e., the bits describing excitations  $b$ and $c$ are entangled with each other. But the bits in the hole are all supposed to be entangled with the radiation, so they {\it cannot} entangle with each other (this follows from the `monogamy of entanglement'). Thus, Susskind's complementarity is not possible.

It had never been clear how Susskind's postulates could be realized, so perhaps there was no great surprise that complementarity could be ruled out. But at first it appeared that the AMPS argument would also rule out the possibility of fuzzball complementarity. The AMPS argument focused on the Hawking modes around the horizon, which have energies $E\sim T$. But if the region around the horizon cannot be a vacuum, then won't every infalling observer interact and scatter off whatever degrees of freedom are present at the fuzzball surface? In that case, it seemed that one could not have fuzzball complementarity either.

In 2014, it was found that there is a flaw in the AMPS argument: while the AMPS argument does rule out Susskind's complementarity, it cannot rule out fuzzball complementarity \cite{flaw}. To see this flaw, let us recall the argument in more detail.

At first sight, the goal of the AMPS argument seems puzzling. Hawking had shown that if the region around the horizon was a vacuum, then the entanglement of the radiation with the hole would keep growing monotonically. AMPS required that the entanglement should {\it not} keep rising, and concluded that the region around the horizon {\it cannot} be a vacuum. Is the AMPS statement not the same as Hawking's statement? Even the line of argumentation was similar: AMPS considered the entangled pairs used by Hawking and described them by the same bit model that was used in the small corrections theorem to make Hawking's argument rigorous.

In fact, AMPS {\it did} try to do something a little different from Hawking, by making an extra assumption. It is this extra assumption that provides the loophole allowing fuzzball complementarity to exist, so let us recall the assumptions of AMPS in a little more detail. 

AMPS required that (i) The entanglement of the radiation with the hole should come down to zero by the end of evaporation, just like it does for  a normal body. (ii) Outside a surface at $r=2GM+\epsilon$, physics should be `normal'; i.e., we should just get the usual low energy effective field theory. (iii) The surface at $r=2GM+\epsilon$ responds to infalling objects in a causal manner; i.e., it cannot alter its state in response to an infalling particle {\it before} the  infalling particle reaches this surface.

We can now see the difference between what AMPS and Hawking could argue from their respective assumptions. Hawking's argument can indeed be recast as saying that if the entanglement has to come down, then the dynamics around the horizon cannot be that of a vacuum region. But the radiated quanta have a long wavelength $\lambda \sim 1/T\sim 2GM$, so the required departure from semiclassical evolution  can be a sharp departure at some location outside the horizon, or a gentle one, say over the region $2GM\lesssim r\lesssim 4GM$. AMPS  {\it postulated} that physics is normal at $r>2GM+\epsilon$, so any departure from vacuum dynamics must take place sharply in the region $2GM<r<2GM+\epsilon$. In other words, the vacuum around the horizon must be replaced by a sharp `firewall'.

But instead of focusing on Hawking modes which have $E\sim T$, consider the infall of a  shell with $E\gg T$. Since physics is `normal' at $r>2GM+\epsilon$, the usual dynamics of gravity in this region will create  a {\it new} horizon at
\be
R_h^{new}=2G(M+E)\,,
\label{uthree}
\ee
which is a location outside the stretched horizon at $r=2GM+\epsilon$. Light cones `point inwards' in the region $R_{shell}<r<R_h^{new}$, where $R_{shell}$ is the location of the shell as it falls in.  Thus, radiation from the stretched horizon  cannot escape to infinity, in contradiction to requirement (i). This difficulty would be avoided if the stretched horizon surface jumped out to a location outside $R_h^{new}$ {\it before} the infalling shell reached $r=R_h^{new}$, but this is disallowed by postulate (iii). Thus, we see that there is an internal conflict in the assumptions (i)-(iii) used by AMPS: an infalling shell creates a new horizon that causally traps the radiation trying to emerge from the stretched horizon, so information cannot actually emerge like it does for a normal body.

We can now see how fuzzball dynamics can evade the above situation. Consider a fuzzball of mass $M$, with a surface that is at $2GM+\epsilon$. Let a shell of energy $E$ fall onto this fuzzball. As the shell approaches the radius $R_h^{new}$ given by (\ref{uthree}), the fuzzball structure begins to nucleate just {\it outside} $r=R_h^{new}$. This nucleation is possible because the theory contains extended objects (fuzzballs) of all  radii, and the gravity wavefunctional has the tail indicated in fig.\ref{fig:fuzzywavefunction}(b) over fuzzball configurations of radius $\sim R_h^{new}$. As the shell passes through the radius $r=R_h^{new}$, the virtual fluctuations (vecros) represented by this tail transition to on-shell fuzzballs, with the wavefunctional evolving as in fig.\ref{fig:fuzzywavefunction}(c). 

Let us return to the issue of how  fuzzball complementarity would work.  The degrees of freedom that are analogous to the open strings in AdS/CFT are the {\it new} fuzzballs that are created by the energy of the infalling object -- in the above discussion, this infalling object is the shell of energy $E$. The analogue of the evolution in the CFT would be the continued spreading of the wavefunctional in fig.\ref{fig:fuzzywavefunction}(c) away from the location of the central potential dip. The graviton falling into AdS in AdS/CFT is analogous to the progressive infall of the shell into the interior geometry of the classical hole. 

With this understanding of fuzzball complementarity, we can see how the condition $E\gg T$ bypasses the AMPS argument. The fuzzball of mass $M$ has entropy $S_{bek}[M]$ and therefor a number of possible states
\be
N_{states}[M]\sim e^{S_{bek}[M]}\,.
\ee
After the shell of energy $E$ comes in, the total number of accessible states is
\be
N_{states}[M+E]\sim e^{S_{bek}[M+E]}\,.
\ee
Assuming $E\ll M$ for simplicity, we have 
\be
S_{bek}[M+E]\approx S_{bek}[M]+{dS_{bek}\over dE}[M] E=S_{bek}[M]+{E\over T}\,.
\ee
Thus,
\be
{N_{states}[M+E]\over N_{states}[M]}\sim e^{E\over T}\,.
\ee
Thus, for $E\gg T$ most of the accessible states are {\it new} states, not present in the hole of mass $M$. The AMPS argument against complementarity was based on the assumption  that the states of the hole were, after the halfway evaporation point, maximally entangled with the radiation. {\it But these new states created by the infall    are not entangled with anything.}   The spread of the wavefunctional over these newly accessible states can give the `fuzzball complementarity' description of the shell, where the shell appears to fall through the interior of a classical black hole geometry.

In \cite{model} a bit model was given that exhibits fuzzball complementarity. 

We thus see that fuzzball complementarity is {\it possible}; the AMPS argument cannot establish that infallers with energy $E\gg T$ feel a firewall at the horizon. On the other hand, there is no reason that the dynamics of fuzzballs should be such that we {\it do} get fuzzball complementarity.  While the information paradox is a puzzle that needs to be resolved (otherwise we violate quantum theory), there is no a priori requirement on what an infalling observer should feel. In other words, it is completely consistent that the infalling observer smashes onto the surface of a fuzzball, with no feeling on continuing to fall smoothly through in any choice of variables. However, the AMPS argument cannot establish that the infaller feels a firewall,  since for an infaller with $E\gg T$ most of the accessible degrees of freedom are not entangled with anything, even if the hole is old (i.e, past the halfway evaporation point).

\subsection{Wormholes}

In 2013, Maldacena and Susskind  \cite{cool} proposed a picture of the hole where the horizon had the traditional semiclassical form, but tiny wormholes linked the inside and outside members of the entangled pairs $(b,c)$ of (\ref{utwo}). This proposal led to much confusion. Were the wormholes just an alternative depiction of the fact that the two members of the Hawking pair were entangled? If so, how does wormhole picture say anything new about the entanglement problem? Or were the wormholes indicative of a nonlocal Hamiltonian interaction between the region inside the hole and the radiated quanta near infinity? These two regions are causally separated from each other, and there is no evidence of such large scale violation of causality in string theory.

While the wormhole proposal remained unclear, it led to a belief about black holes which turned out to be incorrect. This belief was as follows. Suppose we accept that the black hole radiates from its surface like a normal body. Fuzzballs give such a behavior for the hole, so we can assume that the black hole is a fuzzball in its exact quantum gravity description. But, according to the above-mentioned  belief, there is a `code subspace' out of all the complicated quantum gravitational degrees of freedom which describes the low energy dynamics of the fuzzball, and this low energy dynamics is to a first approximation the usual semiclassical dynamics of the traditional hole.

However, the Effective small corrections theorem \cite{contrasting} showed that such a code subspace cannot exist. This theorem is a small extension of the small corrections theorem, and says the following. Consider the exact quantum gravity Hilbert space of the black hole region, which we can take, for concreteness, as the region $r<4GM$. Suppose there is a linear subspace of this exact Hilbert space (the code subspace) in which the evolution approximates the usual semiclassical evolution. (We allow the possibility that this code subspace gives the semiclassical evolution for only a few crossing times; after that we may need a different subspace to function as the code subspace.) Finally, we assume that there are no nonlocal Hamiltonian interactions between the hole and the region near infinity. Then the theorem says that with these assumptions, the entanglement $S_{ent}$ in the {\it exact} theory will have to keep rising monotonically. In other words, if the Page curve is to come down in the exact theory, then we cannot have a code subspace which gives semiclassical dynamics to leading order.

The intuitive idea behind the Effective small corrections theorem is simple. If a semiclassical horizon emerges as a leading order approximation to the exact dynamics, then Hawking pairs will be created in this semiclassical description, with the form (\ref{utwo}). Then the entanglement $S_{ent}$  of the hole with the radiation will keep rising monotonically in this semiclassical approximation. The small corrections theorem can then be used to prove that the entanglement  will keep rising monotonically in the {\it exact} theory, since evolution in the code subspace of the exact theory departs only in a `small' way from semiclassical evolution.

There were attempts to evade this conclusion by making the code subspace using the bits at infinity along with the bits in the black hole region (the so-called $A=R_B$ models), but in this case one does not obtain semiclassical dynamics around the horizon because the dynamics of the bits at infinity is not the one required by semiclassical dynamics in the horizon region.

\section{Summary}

Hawking's black hole information paradox posed a remarkable challenge to the unification of general relativity and quantum theory. General relativity predicts that black holes will form, and the evaporation of these holes appears to violate the unitarity of quantum theory. The resolution of the puzzle through string theory is equally remarkable: quantum gravity effects modify the entire region inside the horizon, so that we get a fuzzball ---  a ball of stringy matter with no horizon. This fuzzball radiates from its surface like a normal body; i.e., not by the creation of pairs from the vacuum. Thus, there is no information paradox.

Some people found this resolution of the puzzle  surprising, because one usually assumes that quantum gravity effects will only stretch over distances  $\sim l_p$. After all, Planck length  is the only length scale that we can make from the fundamental constants $c, \hbar, G$.  But a large black hole is made of a large number of elementary quanta $N$, so we need to ask if the size of a bound state is $\sim l_p$ or $\sim N^\alpha l_p$ for some $\alpha>0$. The fuzzball constructions indicate that the latter is true, with $\alpha$ such that the radius of the generic bound state is order the horizon radius.

The fuzzball paradigm emerged soon after the discovery of D-branes and the construction of black holes as bound states of branes. In \cite{emission} an estimate of the radius of  brane bound states gave a result of order of the horizon radius. Over the years several families of brane bound states have been explicitly constructed, and in each case we find a `fuzzball'; i.e., an object with no horizon. Note that a fuzzball is in general a very quantum object, though we often construct examples of fuzzballs in terms of a classical geometry.  An analogy is black body radiation in a box: the general state is very far from classical, with occupation number $\langle n_{\vec k}\rangle \sim O(1)$ for the typical mode $\vec k$. But to get an estimate of the energy or pressure of this radiation, we can look at specific states where we place all the energy in a few harmonics with $\langle n_{\vec k}\rangle \gg 1$; such states can be understood in terms of the classical electromagnetic fields $\vec E, \vec B$.  The classical fuzzball geometries are the analogues of these latter states, and help us to understand how the no-hair theorems can be broken in string theory to yield horizon-sized structures that describe the microstates of the  black hole. But the relevant aspect of all fuzzball solutions -- classical or quantum -- is that there is no horizon around which the usual low energy semiclassical evolution holds.

Since all the microstates constructed so far have turned out to be fuzzballs,  it does not appear plausible that the remaining microstates do have a horizon and that the information puzzle  is resolved in string theory by some alternative to the fuzzball paradigm. Nevertheless, a lot of confusion has persisted over the years due to attempts where one tries to maintain the traditional semiclassical horizon and yet get the information to emerge in the Hawking radiation. Some examples of such attempts are the following:

\b

(i) The small corrections idea was based on assuming  a  semiclassical horizon  at leading order, and arguing that `subleading saddle points' in the Euclidean path integral would introduce delicate correlations in the radiation to remove the problematic entanglement \cite{hawkingretract}. The `Small corrections theorem' showed that this idea cannot work \cite{cern}.

(ii) The idea of complementarity postulated that with one choice of variables the information reflects off a stretched horizon outside the hole, while with another choice we get the usual semiclassical vacuum around the horizon. The AMPS firewall  argument \cite{amps} showed that such complementarity is not possible.

(iii)  Once it became clear that microstates were fuzzballs, there were attempts to argue that the low energy dynamics of fuzzballs would be captured by evolution in a `code subspace' of the full Hilbert space, and that  dynamics in this code subspace would be approximately the traditional  semiclassical dynamics around a horizon. The `Effective small corrections theorem' (an extension of the small corrections theorem) showed  that this is not possible \cite{contrasting}.

\b

In short, the situation in string theory is the following. There is only {\it one} exact description of a black hole microstate; this state is in general a very messy quantum fuzzball, though for special states it may be well approximated by a classical geometry without horizon. The fuzzball radiates from its surface like any other quantum body, through quanta of energy, $E\sim T$ where $T$ is the temperature of the hole. The dynamics of these radiated quanta near the fuzzball surface depends on the details of the fuzzball microstate; that is, how these quanta are able to carry the information of the hole. There is {\it no} description of the microstate in which we obtain the semiclassical horizon dynamics for these $E\sim T$ quanta. The detailed arguments for this are covered in the results (i)-(iii) above, but in essence the reason is simple: if one could replace the fuzzball by an effective semiclassical horizon, then these $E\sim T$ modes will be produced as members of entangled Hawking pairs, and will not carry any information since the vacuum does not have any information about the microstate. 

The picture of the vacuum described by the vecro hypothesis shows how the semiclassical approximation can break down in gravitational collapse and create fuzzballs,  even though the curvature is low at the horizon. The key is that in this situation, space is stretching too {\it fast}: light cannot travel across the stretching region to create the correlations required to obtain a vacuum on the stretched manifold. We noted that in other low curvature situations like gravitational waves or star formation, we do not have this fast stretching, and so the semiclassical approximation does not break down.  At the scale of the cosmological horizon we again encounter fast stretching, and here we do find the need for unexplained new sources of energy, like dark energy and the Early Dark Energy that might explain the Hubble tension. 

One can extract the essential ideas of string theory in an abstract
form \cite{Mathur:2024ncp}; it is plausible that these ideas can be incorporated into other
alternatives for quantizing gravity. In particular, the vecro picture
of the gravitational vacuum can be a starting point for how to think
of spacetime in any theory of gravity; this picture is essential to
resolving the information paradox while preserving causality and
locality.

\section*{Acknowledgments}

We are grateful to all the people who have worked over the years on the black hole information puzzle, and in particular to all those who have contributed to the fuzzball paradigm. This  work was supported in part by DOE grant DE-FG02-91ER-40690.

\bibliographystyle{utphys}
\bibliography{chapter}

\providecommand{\href}[2]{#2}\begingroup\raggedright\begin{thebibliography}{10}

\bibitem{bek}
J.~D. Bekenstein, ``{Black holes and entropy},'' \href{http://dx.doi.org/10.1103/PhysRevD.7.2333}{{\em Phys. Rev. D} {\bfseries 7} (1973) 2333--2346}.

\bibitem{hawking1}
S.~W. Hawking, ``{Particle Creation by Black Holes},'' \href{http://dx.doi.org/10.1007/BF02345020}{{\em Commun. Math. Phys.} {\bfseries 43} (1975) 199--220}. [Erratum: Commun.Math.Phys. 46, 206 (1976)].

\bibitem{hawking2}
S.~W. Hawking, ``Breakdown of predictability in gravitational collapse,'' \href{http://dx.doi.org/10.1103/PhysRevD.14.2460}{{\em Phys. Rev. D} {\bfseries 14} (Nov, 1976) 2460--2473}. \url{https://link.aps.org/doi/10.1103/PhysRevD.14.2460}.

\bibitem{cern}
S.~D. Mathur, ``{The Information paradox: A Pedagogical introduction},'' \href{http://dx.doi.org/10.1088/0264-9381/26/22/224001}{{\em Class. Quant. Grav.} {\bfseries 26} (2009) 224001}, \href{http://arxiv.org/abs/0909.1038}{{\ttfamily arXiv:0909.1038 [hep-th]}}.

\bibitem{wald}
R.~M. Wald, ``Black hole entropy is the noether charge,'' \href{http://dx.doi.org/10.1103/PhysRevD.48.R3427}{{\em Phys. Rev. D} {\bfseries 48} (Oct, 1993) R3427--R3431}. \url{https://link.aps.org/doi/10.1103/PhysRevD.48.R3427}.

\bibitem{dabholkar}
A.~Dabholkar, ``{Exact counting of black hole microstates},'' \href{http://dx.doi.org/10.1103/PhysRevLett.94.241301}{{\em Phys. Rev. Lett.} {\bfseries 94} (2005) 241301}, \href{http://arxiv.org/abs/hep-th/0409148}{{\ttfamily arXiv:hep-th/0409148}}.

\bibitem{sen1}
A.~Sen, ``{Black hole solutions in heterotic string theory on a torus},'' \href{http://dx.doi.org/10.1016/0550-3213(95)00063-X}{{\em Nucl. Phys. B} {\bfseries 440} (1995) 421--440}, \href{http://arxiv.org/abs/hep-th/9411187}{{\ttfamily arXiv:hep-th/9411187}}.

\bibitem{sen2}
A.~Sen, ``{Extremal black holes and elementary string states},'' \href{http://dx.doi.org/10.1142/S0217732395002234}{{\em Mod. Phys. Lett. A} {\bfseries 10} (1995) 2081--2094}, \href{http://arxiv.org/abs/hep-th/9504147}{{\ttfamily arXiv:hep-th/9504147}}.

\bibitem{sv}
A.~Strominger and C.~Vafa, ``{Microscopic origin of the Bekenstein-Hawking entropy},'' \href{http://dx.doi.org/10.1016/0370-2693(96)00345-0}{{\em Phys. Lett. B} {\bfseries 379} (1996) 99--104}, \href{http://arxiv.org/abs/hep-th/9601029}{{\ttfamily arXiv:hep-th/9601029}}.

\bibitem{callanmalda}
C.~G. Callan and J.~M. Maldacena, ``{D-brane approach to black hole quantum mechanics},'' \href{http://dx.doi.org/10.1016/0550-3213(96)00225-8}{{\em Nucl. Phys. B} {\bfseries 472} (1996) 591--610}, \href{http://arxiv.org/abs/hep-th/9602043}{{\ttfamily arXiv:hep-th/9602043}}.

\bibitem{hms}
G.~T. Horowitz, J.~M. Maldacena, and A.~Strominger, ``{Nonextremal black hole microstates and U duality},'' \href{http://dx.doi.org/10.1016/0370-2693(96)00738-1}{{\em Phys. Lett. B} {\bfseries 383} (1996) 151--159}, \href{http://arxiv.org/abs/hep-th/9603109}{{\ttfamily arXiv:hep-th/9603109}}.

\bibitem{dasmathur}
S.~R. Das and S.~D. Mathur, ``{Comparing decay rates for black holes and D-branes},'' \href{http://dx.doi.org/10.1016/0550-3213(96)00453-1}{{\em Nucl. Phys. B} {\bfseries 478} (1996) 561--576}, \href{http://arxiv.org/abs/hep-th/9606185}{{\ttfamily arXiv:hep-th/9606185}}.

\bibitem{maldastrom}
J.~Maldacena and A.~Strominger, ``Black hole greybody factors and d-brane spectroscopy,'' \href{http://dx.doi.org/10.1103/PhysRevD.55.861}{{\em Phys. Rev. D} {\bfseries 55} (Jan, 1997) 861--870}. \url{https://link.aps.org/doi/10.1103/PhysRevD.55.861}.

\bibitem{lm4}
O.~Lunin and S.~D. Mathur, ``{AdS / CFT duality and the black hole information paradox},'' \href{http://dx.doi.org/10.1016/S0550-3213(01)00620-4}{{\em Nucl. Phys. B} {\bfseries 623} (2002) 342--394}, \href{http://arxiv.org/abs/hep-th/0109154}{{\ttfamily arXiv:hep-th/0109154}}.

\bibitem{lmm}
O.~Lunin, J.~M. Maldacena, and L.~Maoz, ``{Gravity solutions for the D1-D5 system with angular momentum},'' \href{http://arxiv.org/abs/hep-th/0212210}{{\ttfamily arXiv:hep-th/0212210}}.

\bibitem{skenderis}
I.~Kanitscheider, K.~Skenderis, and M.~Taylor, ``{Fuzzballs with internal excitations},'' \href{http://dx.doi.org/10.1088/1126-6708/2007/06/056}{{\em JHEP} {\bfseries 06} (2007) 056}, \href{http://arxiv.org/abs/0704.0690}{{\ttfamily arXiv:0704.0690 [hep-th]}}.

\bibitem{bdkr}
V.~Balasubramanian, J.~de~Boer, E.~Keski-Vakkuri, and S.~F. Ross, ``{Supersymmetric conical defects: Towards a string theoretic description of black hole formation},'' \href{http://dx.doi.org/10.1103/PhysRevD.64.064011}{{\em Phys. Rev. D} {\bfseries 64} (2001) 064011}, \href{http://arxiv.org/abs/hep-th/0011217}{{\ttfamily arXiv:hep-th/0011217}}.

\bibitem{mm}
J.~M. Maldacena and L.~Maoz, ``{Desingularization by rotation},'' \href{http://dx.doi.org/10.1088/1126-6708/2002/12/055}{{\em JHEP} {\bfseries 12} (2002) 055}, \href{http://arxiv.org/abs/hep-th/0012025}{{\ttfamily arXiv:hep-th/0012025}}.

\bibitem{multiwound}
O.~Lunin and S.~D. Mathur, ``{Metric of the multiply wound rotating string},'' \href{http://dx.doi.org/10.1016/S0550-3213(01)00321-2}{{\em Nucl. Phys. B} {\bfseries 610} (2001) 49--76}, \href{http://arxiv.org/abs/hep-th/0105136}{{\ttfamily arXiv:hep-th/0105136}}.

\bibitem{martinec}
E.~J. Martinec and W.~McElgin, ``{String theory on AdS orbifolds},'' \href{http://dx.doi.org/10.1088/1126-6708/2002/04/029}{{\em JHEP} {\bfseries 04} (2002) 029}, \href{http://arxiv.org/abs/hep-th/0106171}{{\ttfamily arXiv:hep-th/0106171}}.

\bibitem{supertubes}
R.~Emparan, D.~Mateos, and P.~K. Townsend, ``{Supergravity supertubes},'' \href{http://dx.doi.org/10.1088/1126-6708/2001/07/011}{{\em JHEP} {\bfseries 07} (2001) 011}, \href{http://arxiv.org/abs/hep-th/0106012}{{\ttfamily arXiv:hep-th/0106012}}.

\bibitem{lm5}
O.~Lunin and S.~D. Mathur, ``{Statistical interpretation of Bekenstein entropy for systems with a stretched horizon},'' \href{http://dx.doi.org/10.1103/PhysRevLett.88.211303}{{\em Phys. Rev. Lett.} {\bfseries 88} (2002) 211303}, \href{http://arxiv.org/abs/hep-th/0202072}{{\ttfamily arXiv:hep-th/0202072}}.

\bibitem{phase}
S.~D. Mathur, ``{Black hole size and phase space volumes},'' \href{http://arxiv.org/abs/0706.3884}{{\ttfamily arXiv:0706.3884 [hep-th]}}.

\bibitem{rychkov}
V.~S. Rychkov, ``{D1-D5 black hole microstate counting from supergravity},'' \href{http://dx.doi.org/10.1088/1126-6708/2006/01/063}{{\em JHEP} {\bfseries 01} (2006) 063}, \href{http://arxiv.org/abs/hep-th/0512053}{{\ttfamily arXiv:hep-th/0512053}}.

\bibitem{mss}
S.~D. Mathur, A.~Saxena, and Y.~K. Srivastava, ``{Constructing `hair' for the three charge hole},'' \href{http://dx.doi.org/10.1016/j.nuclphysb.2003.12.022}{{\em Nucl. Phys. B} {\bfseries 680} (2004) 415--449}, \href{http://arxiv.org/abs/hep-th/0311092}{{\ttfamily arXiv:hep-th/0311092}}.

\bibitem{lunin}
O.~Lunin, ``{Adding momentum to D-1 - D-5 system},'' \href{http://dx.doi.org/10.1088/1126-6708/2004/04/054}{{\em JHEP} {\bfseries 04} (2004) 054}, \href{http://arxiv.org/abs/hep-th/0404006}{{\ttfamily arXiv:hep-th/0404006}}.

\bibitem{gms1}
S.~Giusto, S.~D. Mathur, and A.~Saxena, ``{Dual geometries for a set of 3-charge microstates},'' \href{http://dx.doi.org/10.1016/j.nuclphysb.2004.09.001}{{\em Nucl. Phys. B} {\bfseries 701} (2004) 357--379}, \href{http://arxiv.org/abs/hep-th/0405017}{{\ttfamily arXiv:hep-th/0405017}}.

\bibitem{gms2}
S.~Giusto, S.~D. Mathur, and A.~Saxena, ``{3-charge geometries and their CFT duals},'' \href{http://dx.doi.org/10.1016/j.nuclphysb.2005.01.009}{{\em Nucl. Phys. B} {\bfseries 710} (2005) 425--463}, \href{http://arxiv.org/abs/hep-th/0406103}{{\ttfamily arXiv:hep-th/0406103}}.

\bibitem{russo}
A.~Galliani, S.~Giusto, E.~Moscato, and R.~Russo, ``{Correlators at large c without information loss},'' \href{http://dx.doi.org/10.1007/JHEP09(2016)065}{{\em JHEP} {\bfseries 09} (2016) 065}, \href{http://arxiv.org/abs/1606.01119}{{\ttfamily arXiv:1606.01119 [hep-th]}}.

\bibitem{gauntlett}
J.~B. Gutowski, D.~Martelli, and H.~S. Reall, ``{All Supersymmetric solutions of minimal supergravity in six- dimensions},'' \href{http://dx.doi.org/10.1088/0264-9381/20/23/008}{{\em Class. Quant. Grav.} {\bfseries 20} (2003) 5049--5078}, \href{http://arxiv.org/abs/hep-th/0306235}{{\ttfamily arXiv:hep-th/0306235}}.

\bibitem{giustomathur}
S.~Giusto and S.~D. Mathur, ``{Geometry of D1-D5-P bound states},'' \href{http://dx.doi.org/10.1016/j.nuclphysb.2005.09.037}{{\em Nucl. Phys. B} {\bfseries 729} (2005) 203--220}, \href{http://arxiv.org/abs/hep-th/0409067}{{\ttfamily arXiv:hep-th/0409067}}.

\bibitem{bw01}
I.~Bena and N.~P. Warner, ``{One ring to rule them all ... and in the darkness bind them?},'' \href{http://dx.doi.org/10.4310/ATMP.2005.v9.n5.a1}{{\em Adv. Theor. Math. Phys.} {\bfseries 9} no.~5, (2005) 667--701}, \href{http://arxiv.org/abs/hep-th/0408106}{{\ttfamily arXiv:hep-th/0408106}}.

\bibitem{berglund}
P.~Berglund, E.~G. Gimon, and T.~S. Levi, ``{Supergravity microstates for BPS black holes and black rings},'' \href{http://dx.doi.org/10.1088/1126-6708/2006/06/007}{{\em JHEP} {\bfseries 06} (2006) 007}, \href{http://arxiv.org/abs/hep-th/0505167}{{\ttfamily arXiv:hep-th/0505167}}.

\bibitem{bw02}
I.~Bena and N.~P. Warner, ``Bubbling supertubes and foaming black holes,'' \href{http://dx.doi.org/10.1103/PhysRevD.74.066001}{{\em Phys. Rev. D} {\bfseries 74} (Sep, 2006) 066001}. \url{https://link.aps.org/doi/10.1103/PhysRevD.74.066001}.

\bibitem{bwreview}
I.~Bena and N.~P. Warner, ``{Black holes, black rings and their microstates},'' \href{http://dx.doi.org/10.1007/978-3-540-79523-0_1}{{\em Lect. Notes Phys.} {\bfseries 755} (2008) 1--92}, \href{http://arxiv.org/abs/hep-th/0701216}{{\ttfamily arXiv:hep-th/0701216}}.

\bibitem{bw1}
I.~Bena, S.~Giusto, R.~Russo, M.~Shigemori, and N.~P. Warner, ``{Habemus Superstratum! A constructive proof of the existence of superstrata},'' \href{http://dx.doi.org/10.1007/JHEP05(2015)110}{{\em JHEP} {\bfseries 05} (2015) 110}, \href{http://arxiv.org/abs/1503.01463}{{\ttfamily arXiv:1503.01463 [hep-th]}}.

\bibitem{bw2}
I.~Bena, S.~Giusto, E.~J. Martinec, R.~Russo, M.~Shigemori, D.~Turton, and N.~P. Warner, ``{Smooth horizonless geometries deep inside the black-hole regime},'' \href{http://dx.doi.org/10.1103/PhysRevLett.117.201601}{{\em Phys. Rev. Lett.} {\bfseries 117} no.~20, (2016) 201601}, \href{http://arxiv.org/abs/1607.03908}{{\ttfamily arXiv:1607.03908 [hep-th]}}.

\bibitem{jmart}
V.~Jejjala, O.~Madden, S.~F. Ross, and G.~Titchener, ``{Non-supersymmetric smooth geometries and D1-D5-P bound states},'' \href{http://dx.doi.org/10.1103/PhysRevD.71.124030}{{\em Phys. Rev. D} {\bfseries 71} (2005) 124030}, \href{http://arxiv.org/abs/hep-th/0504181}{{\ttfamily arXiv:hep-th/0504181}}.

\bibitem{myers}
V.~Cardoso, O.~J.~C. Dias, J.~L. Hovdebo, and R.~C. Myers, ``{Instability of non-supersymmetric smooth geometries},'' \href{http://dx.doi.org/10.1103/PhysRevD.73.064031}{{\em Phys. Rev. D} {\bfseries 73} (2006) 064031}, \href{http://arxiv.org/abs/hep-th/0512277}{{\ttfamily arXiv:hep-th/0512277}}.

\bibitem{cm1}
B.~D. Chowdhury and S.~D. Mathur, ``{Radiation from the non-extremal fuzzball},'' \href{http://dx.doi.org/10.1088/0264-9381/25/13/135005}{{\em Class. Quant. Grav.} {\bfseries 25} (2008) 135005}, \href{http://arxiv.org/abs/0711.4817}{{\ttfamily arXiv:0711.4817 [hep-th]}}.

\bibitem{cm2}
B.~D. Chowdhury and S.~D. Mathur, ``{Pair creation in non-extremal fuzzball geometries},'' \href{http://dx.doi.org/10.1088/0264-9381/25/22/225021}{{\em Class. Quant. Grav.} {\bfseries 25} (2008) 225021}, \href{http://arxiv.org/abs/0806.2309}{{\ttfamily arXiv:0806.2309 [hep-th]}}.

\bibitem{virmani}
B.~Chakrabarty, D.~Turton, and A.~Virmani, ``{Holographic description of non-supersymmetric orbifolded D1-D5-P solutions},'' \href{http://dx.doi.org/10.1007/JHEP11(2015)063}{{\em JHEP} {\bfseries 11} (2015) 063}, \href{http://arxiv.org/abs/1508.01231}{{\ttfamily arXiv:1508.01231 [hep-th]}}.

\bibitem{rotating}
S.~D. Mathur and D.~Turton, ``{Oscillating supertubes and neutral rotating black hole microstates},'' \href{http://dx.doi.org/10.1007/JHEP04(2014)072}{{\em JHEP} {\bfseries 04} (2014) 072}, \href{http://arxiv.org/abs/1310.1354}{{\ttfamily arXiv:1310.1354 [hep-th]}}.

\bibitem{bolts}
G.~Bossard, S.~Katmadas, and D.~Turton, ``{Two Kissing Bolts},'' \href{http://dx.doi.org/10.1007/JHEP02(2018)008}{{\em JHEP} {\bfseries 02} (2018) 008}, \href{http://arxiv.org/abs/1711.04784}{{\ttfamily arXiv:1711.04784 [hep-th]}}.

\bibitem{nonex}
B.~Ganchev, S.~Giusto, A.~Houppe, R.~Russo, and N.~P. Warner, ``{Microstrata},'' \href{http://dx.doi.org/10.1007/JHEP10(2023)163}{{\em JHEP} {\bfseries 10} (2023) 163}, \href{http://arxiv.org/abs/2307.13021}{{\ttfamily arXiv:2307.13021 [hep-th]}}.

\bibitem{ghm}
B.~Guo, S.~Hampton, and S.~D. Mathur, ``{Can we observe fuzzballs or firewalls?},'' \href{http://dx.doi.org/10.1007/JHEP07(2018)162}{{\em JHEP} {\bfseries 07} (2018) 162}, \href{http://arxiv.org/abs/1711.01617}{{\ttfamily arXiv:1711.01617 [hep-th]}}.

\bibitem{bubble}
E.~Witten, ``{Instability of the Kaluza-Klein Vacuum},'' \href{http://dx.doi.org/10.1016/0550-3213(82)90007-4}{{\em Nucl. Phys. B} {\bfseries 195} (1982) 481--492}.

\bibitem{what}
S.~D. Mathur, ``{What prevents gravitational collapse in string theory?},'' \href{http://dx.doi.org/10.1142/S0218271816440181}{{\em Int. J. Mod. Phys. D} {\bfseries 25} no.~12, (2016) 1644018}, \href{http://arxiv.org/abs/1609.05222}{{\ttfamily arXiv:1609.05222 [hep-th]}}.

\bibitem{gibbonswarner}
G.~W. Gibbons and N.~P. Warner, ``{Global structure of five-dimensional fuzzballs},'' \href{http://dx.doi.org/10.1088/0264-9381/31/2/025016}{{\em Class. Quant. Grav.} {\bfseries 31} (2014) 025016}, \href{http://arxiv.org/abs/1305.0957}{{\ttfamily arXiv:1305.0957 [hep-th]}}.

\bibitem{heidmann}
I.~Bah and P.~Heidmann, ``{Topological stars, black holes and generalized charged Weyl solutions},'' \href{http://dx.doi.org/10.1007/JHEP09(2021)147}{{\em JHEP} {\bfseries 09} (2021) 147}, \href{http://arxiv.org/abs/2012.13407}{{\ttfamily arXiv:2012.13407 [hep-th]}}.

\bibitem{bala}
V.~Balasubramanian, E.~G. Gimon, and T.~S. Levi, ``{Four Dimensional Black Hole Microstates: From D-branes to Spacetime Foam},'' \href{http://dx.doi.org/10.1088/1126-6708/2008/01/056}{{\em JHEP} {\bfseries 01} (2008) 056}, \href{http://arxiv.org/abs/hep-th/0606118}{{\ttfamily arXiv:hep-th/0606118}}.

\bibitem{ecometric}
S.~Toktarbay and H.~Quevedo, ``{A stationary q-metric},'' \href{http://dx.doi.org/10.1134/S0202289314040136}{{\em Grav. Cosmol.} {\bfseries 20} (2014) 252}, \href{http://arxiv.org/abs/1510.04155}{{\ttfamily arXiv:1510.04155 [gr-qc]}}.

\bibitem{bkl}
V.~A. Belinskii, E.~M. Lifshitz, and I.~M. Khalatnikov, ``{On a general cosmological solution of the einstein equations with a time singularity},'' {\em Zh. Eksp. Teor. Fiz.} {\bfseries 62} (1972) 1606--1613.

\bibitem{eco1}
S.~D. Mathur and M.~Mehta, ``{The universality of black hole thermodynamics},'' \href{http://dx.doi.org/10.1142/S0218271823410031}{{\em Int. J. Mod. Phys. D} {\bfseries 32} no.~14, (2023) 2341003}, \href{http://arxiv.org/abs/2305.12003}{{\ttfamily arXiv:2305.12003 [hep-th]}}.

\bibitem{eco2}
S.~D. Mathur and M.~Mehta, ``{The universal thermodynamic properties of extremely compact objects},'' \href{http://dx.doi.org/10.1088/1361-6382/ad869e}{{\em Class. Quant. Grav.} {\bfseries 41} no.~23, (2024) 235011}, \href{http://arxiv.org/abs/2402.13166}{{\ttfamily arXiv:2402.13166 [hep-th]}}.

\bibitem{tunnel}
S.~D. Mathur, ``{Tunneling into fuzzball states},'' \href{http://dx.doi.org/10.1007/s10714-009-0837-3}{{\em Gen. Rel. Grav.} {\bfseries 42} (2010) 113--118}, \href{http://arxiv.org/abs/0805.3716}{{\ttfamily arXiv:0805.3716 [hep-th]}}.

\bibitem{kraus}
P.~Kraus and S.~D. Mathur, ``{Nature abhors a horizon},'' \href{http://dx.doi.org/10.1142/S0218271815430038}{{\em Int. J. Mod. Phys. D} {\bfseries 24} no.~12, (2015) 1543003}, \href{http://arxiv.org/abs/1505.05078}{{\ttfamily arXiv:1505.05078 [hep-th]}}.

\bibitem{puhm}
I.~Bena, D.~R. Mayerson, A.~Puhm, and B.~Vercnocke, ``{Tunneling into Microstate Geometries: Quantum Effects Stop Gravitational Collapse},'' \href{http://dx.doi.org/10.1007/JHEP07(2016)031}{{\em JHEP} {\bfseries 07} (2016) 031}, \href{http://arxiv.org/abs/1512.05376}{{\ttfamily arXiv:1512.05376 [hep-th]}}.

\bibitem{vecro}
S.~D. Mathur, ``{The VECRO hypothesis},'' \href{http://arxiv.org/abs/2001.11057}{{\ttfamily arXiv:2001.11057 [hep-th]}}.

\bibitem{elastic}
S.~D. Mathur, ``{The elastic vacuum},'' \href{http://dx.doi.org/10.1142/S0218271821410017}{{\em Int. J. Mod. Phys. D} {\bfseries 30} no.~14, (2021) 2141001}, \href{http://arxiv.org/abs/2105.06963}{{\ttfamily arXiv:2105.06963 [hep-th]}}.

\bibitem{secret}
S.~D. Mathur, ``{The secret structure of the gravitational vacuum},'' \href{http://arxiv.org/abs/2405.08945}{{\ttfamily arXiv:2405.08945 [hep-th]}}.

\bibitem{thooft}
C.~R. Stephens, G.~'t~Hooft, and B.~F. Whiting, ``{Black hole evaporation without information loss},'' \href{http://dx.doi.org/10.1088/0264-9381/11/3/014}{{\em Class. Quant. Grav.} {\bfseries 11} (1994) 621--648}, \href{http://arxiv.org/abs/gr-qc/9310006}{{\ttfamily arXiv:gr-qc/9310006}}.

\bibitem{susskindetal}
L.~Susskind, L.~Thorlacius, and J.~Uglum, ``{The Stretched horizon and black hole complementarity},'' \href{http://dx.doi.org/10.1103/PhysRevD.48.3743}{{\em Phys. Rev. D} {\bfseries 48} (1993) 3743--3761}, \href{http://arxiv.org/abs/hep-th/9306069}{{\ttfamily arXiv:hep-th/9306069}}.

\bibitem{emission}
S.~D. Mathur, ``{Emission rates, the correspondence principle and the information paradox},'' \href{http://dx.doi.org/10.1016/S0550-3213(98)00336-8}{{\em Nucl. Phys. B} {\bfseries 529} (1998) 295--320}, \href{http://arxiv.org/abs/hep-th/9706151}{{\ttfamily arXiv:hep-th/9706151}}.

\bibitem{fuzzballreview}
S.~D. Mathur, ``{The Fuzzball proposal for black holes: An Elementary review},'' \href{http://dx.doi.org/10.1002/prop.200410203}{{\em Fortsch. Phys.} {\bfseries 53} (2005) 793--827}, \href{http://arxiv.org/abs/hep-th/0502050}{{\ttfamily arXiv:hep-th/0502050}}.

\bibitem{borunreview}
B.~D. Chowdhury and A.~Virmani, ``{Modave Lectures on Fuzzballs and Emission from the D1-D5 System},'' in {\em {5th Modave Summer School in Mathematical Physics}}.
\newblock 1, 2010.
\newblock \href{http://arxiv.org/abs/1001.1444}{{\ttfamily arXiv:1001.1444 [hep-th]}}.

\bibitem{adscft}
J.~M. Maldacena, ``{The Large N limit of superconformal field theories and supergravity},'' \href{http://dx.doi.org/10.4310/ATMP.1998.v2.n2.a1}{{\em Adv. Theor. Math. Phys.} {\bfseries 2} (1998) 231--252}, \href{http://arxiv.org/abs/hep-th/9711200}{{\ttfamily arXiv:hep-th/9711200}}.

\bibitem{gkp}
S.~S. Gubser, I.~R. Klebanov, and A.~M. Polyakov, ``{Gauge theory correlators from noncritical string theory},'' \href{http://dx.doi.org/10.1016/S0370-2693(98)00377-3}{{\em Phys. Lett. B} {\bfseries 428} (1998) 105--114}, \href{http://arxiv.org/abs/hep-th/9802109}{{\ttfamily arXiv:hep-th/9802109}}.

\bibitem{wittenads}
E.~Witten, ``{Anti-de Sitter space and holography},'' \href{http://dx.doi.org/10.4310/ATMP.1998.v2.n2.a2}{{\em Adv. Theor. Math. Phys.} {\bfseries 2} (1998) 253--291}, \href{http://arxiv.org/abs/hep-th/9802150}{{\ttfamily arXiv:hep-th/9802150}}.

\bibitem{hawkingretract}
S.~W. Hawking, ``{Information loss in black holes},'' \href{http://dx.doi.org/10.1103/PhysRevD.72.084013}{{\em Phys. Rev. D} {\bfseries 72} (2005) 084013}, \href{http://arxiv.org/abs/hep-th/0507171}{{\ttfamily arXiv:hep-th/0507171}}.

\bibitem{amps}
A.~Almheiri, D.~Marolf, J.~Polchinski, and J.~Sully, ``{Black Holes: Complementarity or Firewalls?},'' \href{http://dx.doi.org/10.1007/JHEP02(2013)062}{{\em JHEP} {\bfseries 02} (2013) 062}, \href{http://arxiv.org/abs/1207.3123}{{\ttfamily arXiv:1207.3123 [hep-th]}}.

\bibitem{flaw}
S.~D. Mathur and D.~Turton, ``{The flaw in the firewall argument},'' \href{http://dx.doi.org/10.1016/j.nuclphysb.2014.05.012}{{\em Nucl. Phys. B} {\bfseries 884} (2014) 566--611}, \href{http://arxiv.org/abs/1306.5488}{{\ttfamily arXiv:1306.5488 [hep-th]}}.

\bibitem{model}
S.~D. Mathur, ``{A model with no firewall},'' \href{http://arxiv.org/abs/1506.04342}{{\ttfamily arXiv:1506.04342 [hep-th]}}.

\bibitem{cool}
J.~Maldacena and L.~Susskind, ``{Cool horizons for entangled black holes},'' \href{http://dx.doi.org/10.1002/prop.201300020}{{\em Fortsch. Phys.} {\bfseries 61} (2013) 781--811}, \href{http://arxiv.org/abs/1306.0533}{{\ttfamily arXiv:1306.0533 [hep-th]}}.

\bibitem{contrasting}
B.~Guo, M.~R.~R. Hughes, S.~D. Mathur, and M.~Mehta, ``{Contrasting the fuzzball and wormhole paradigms for black holes},'' \href{http://dx.doi.org/10.3906/fiz-2111-13}{{\em Turk. J. Phys.} {\bfseries 45} no.~6, (2021) 281--365}, \href{http://arxiv.org/abs/2111.05295}{{\ttfamily arXiv:2111.05295 [hep-th]}}.

\bibitem{Mathur:2024ncp}
S.~D. Mathur, ``{How the black hole puzzles are resolved in string theory},'' \href{http://dx.doi.org/10.1007/s10714-024-03336-3}{{\em Gen. Rel. Grav.} {\bfseries 57} no.~1, (2025) 3}.

\end{thebibliography}\endgroup

\end{document}